\documentclass[12pt,preprint,apj,numberedappendix]{emulateapj}

\usepackage{graphicx,apjfonts}
\usepackage{natbib}
\usepackage{epsf}
\usepackage{color}
\shortauthors{Gallazzi et al.}
\shorttitle{Evolution of the ages and metallicities of massive galaxies since $z=0.7$}

\newcommand{\ha}{{\rm H$\alpha$ }}
\newcommand{\hb}{{\rm H$\beta$}}

\newcommand{\oii}{{\rm [O\,{\sc ii}]}}

\newcommand{\oiii}{{\rm [O\,{\sc iii}]}}

\newcommand{\combo}{{\sc combo-17}}

\newcommand{\dn}{$\rm D4000_n$}
\newcommand{\hda}{$\rm H\delta_A$}
\newcommand{\hga}{$\rm H\gamma_A$}
\newcommand{\hbeta}{$\rm H\beta$}
\newcommand{\hdg}{\hbox{H$\delta_A$+H$\gamma_A$}}
\newcommand{\mgfep}{\hbox{$\rm [MgFe]^\prime$}}

\newcommand{\mgtwofe}{\hbox{$\rm [Mg_2Fe]$}}

\def\aj{AJ}%
\def\apj{ApJ}%
\def\apjl{ApJ}%
\def\apjs{ApJS}%
\def\apss{Ap\&SS}%
\def\aap{A\&A}%
\def\aaps{A\&AS}%
\def\mnras{MNRAS}%
\def\pasp{PASP}%
\slugcomment{{\sc ApJ accepted April 21, 2014} }

\begin{document}

\title{Charting the evolution of the ages and metallicities of massive galaxies since $z=0.7$}

\author{Anna Gallazzi\altaffilmark{1,2}, Eric F.\ Bell\altaffilmark{3},
Stefano Zibetti\altaffilmark{1},
Jarle Brinchmann\altaffilmark{4},
Daniel D. Kelson\altaffilmark{5}}

\begin{abstract}
Detailed studies of the stellar populations of intermediate-redshift galaxies can shed light onto the processes responsible
for the growth of the massive galaxy population in the last 8 billion years. We here take a step toward this goal by means
of deep, multi-object rest-frame optical spectroscopy, performed with IMACS on the Magellan telescope, of a sample of $\sim$70 galaxies in
the E-CDFS with redshift $0.65\leq z\leq0.75$, apparent $R>22.7$ $\rm mag_{Vega}$ and stellar mass $>10^{10}M_\odot$. We measure velocity
dispersion and stellar absorption features for individual sources. We interpret them by means of a large Monte Carlo library
of star formation histories, following the Bayesian approach adopted for previous low redshift studies and derive constraints on the
stellar mass, mean stellar age and stellar metallicity of these galaxies. We characterize for the first time the relations between stellar
age and stellar mass and between stellar metallicity and stellar mass at $z\sim0.7$ for the galaxy population as a whole and for quiescent and
star-forming galaxies separately. These relations of increasing age and metallicity with galaxy mass 
for the galaxy population as a whole have a similar shape as the $z\sim0.1$ analog derived for SDSS galaxies, 
but are shifted by $-0.28$~dex in age and by $-0.13$~dex in metallicity, at odds with simple passive evolution.
Considering $z=0.7$ quiescent galaxies alone we find that no additional star formation and chemical enrichment are required for them to
evolve into the present-day quiescent population. However, other observations require that the quiescent population grows from
$z=0.7$ to the present-day. This could be supplied by the quenching of a fraction of $z=0.7$ $M_\star>10^{11}M_\odot$ star-forming galaxies with
metallicities already comparable to those of quiescent galaxies, thus leading to the observed increase of the scatter in age without affecting the metallicity distribution. 
However rapid quenching of the entire population of massive
star-forming galaxies at $z=0.7$ would be inconsistent with the age/metallicity--mass relations for the population as a whole
and with the metallicity distribution of star-forming galaxies only, which are on average 0.12 dex less metal-rich than their local counterparts. This indicates
chemical enrichment until the present in at least a fraction of the $z=0.7$ star-forming galaxies in our sample.
\end{abstract}


\altaffiltext{1} {INAF-Osservatorio Astrofisico di Arcetri, Largo Enrico Fermi 5, 50125 Firenze, 
Italy; \texttt{gallazzi@arcetri.astro.it}} 
\altaffiltext{2} {Dark Cosmology Center, University of Copenhagen, Niels Bohr Institute, 
Juliane Maries Vej 30, 2100 Copenhagen, Denmark}
\altaffiltext{3} {Department of Astronomy, University of Michigan, 500 Church St., Ann Arbor, MI 48109, USA}
\altaffiltext{4} {Leiden Observatory, Leiden University, 2300RA, Leiden, the Netherlands}
\altaffiltext{5} {Observatories of the Carnegie Institution of Washington, Pasadena, CA 91101, USA}

\section{Introduction}\label{sec:intro}
Multi-wavelength and spectroscopic surveys at high redshift have revealed significant evolution from $z\sim1-2$ down to the present-day in the
massive galaxy population. In particular, the star formation rate (SFR) density of the Universe, dominated at all epochs by blue galaxies with
stellar mass in the range $10^{10}-10^{11}M_\odot$ \citep{Panter07,xzz07}, has declined by a factor of $\sim10$ in the last 8 billion years in a
way that is almost independent of galaxy mass \citep[e.g.][]{Hopkins06,smolcic09,karim11,cucciati12}.  This is accompanied by the evolution of
the galaxy stellar mass function indicating that $\sim50$\% of the current stellar mass density  was already in place by $z\sim1$
\citep[e.g.][]{muzzin13}. The massive galaxy population may keep on evolving in number density at $z<1$, albeit at a slower rate than lower-mass
galaxies \citep{moustakas13}. Several studies have shown that the growth of the population as a whole hides a different  evolution of the quiescent and star-forming
galaxy populations. The number density of massive quiescent galaxies grows by $\sim0.2-0.5$~dex since $z\sim1$, while that of star-forming galaxies
remains constant or even declines \citep{muzzin13,moustakas13,Ilbert13,Ilbert10,Brammer11,bell07,faber07,borch06,cimatti06}. The massive end of
the stellar mass function becomes increasingly dominated by quiescent galaxies at lower redshifts \citep{muzzin13,pozzetti10,bell07,bundy06}. By
the present day, quiescent galaxies contain more than half of the mass and metals in stars \citep{bell03,Baldry04,Gallazzi08} and dominate at
stellar masses above $3\times10^{10}M_\odot$.

The physical properties of galaxy stellar populations, such as mean age and metallicity, give us insight into their past history of star formation.
In the local Universe the light-weighted ages and metallicities (stellar and gaseous) of both star-forming and quiescent galaxies have been shown
to correlate to first order with galaxy mass \citep[e.g.][]{gallazzi05,panter08,Tremonti04,mateus06}, with environment playing a stronger role in
less massive galaxies \citep[e.g.][]{bernardi09,Thomas10,pasquali10,pasquali12,Cooper08,Cooper10}. In particular quiescent, elliptical galaxies
follow tight relations, such as the color-magnitude relation and the relation between absorption index strengths and velocity dispersion, originating 
from the increase of their light-weighted age, stellar metallicity and element abundance ratio with galaxy mass
\citep[e.g.][]{trager2000,kuntschner01,Thomas05,nelan05,gallazzi06,graves09a}. This indicates that the stars in present-day more massive
ellipticals have reached a higher degree of chemical enrichment and have formed earlier and on shorter timescales than less massive galaxies. 
The small intrinsic scatter in these observational relations is primarily driven by variation in light-weighted age and, to a smaller extent, stellar
metallicity \citep[e.g.][]{gallazzi06,graves09b}.
Tracing the evolution with redshift of galaxy stellar population properties in relation to
galaxy mass and star formation activity puts additional constraints to the mechanisms leading to the suppression of star formation and the ensuing build-up of the present-day quiescent massive galaxy
population.

The chemical properties of the gas in massive star-forming galaxies have been traced up to $z\sim3$ through emission line analysis
\citep[][]{Erb06,cowie08,Lamareille09,Mannucci10,moustakas11}. These studies have indicated a significant evolution in gas-phase metallicity
with redshift, which is however tied to the evolution in SFR such that star-forming galaxies appear to follow a non-evolving relation between
their mass, gas-phase metallicity and SFR \citep{Mannucci10,LaraLopez10}. On the contrary, stellar population studies are observationally
more challenging as they require deep spectroscopy in order to trace the stellar continuum and the strength of key absorption features
chiefly sensitive to age and metallicity.  Few works have pushed metallicity analysis to $z\sim2$ based on rest-frame
UV absorption features, which trace the {\it youngest} stellar populations \citep{Rix04,Halliday08,Sommariva12}. Studies of the evolution of
the {\it bulk} of stellar populations in galaxies, which require rest-frame optical diagnostics \citep[e.g.][]{worthey94,wo97}, are so far
limited to relatively few deep spectroscopic works at intermediate redshifts. These studies target red-sequence, quiescent galaxies with weak
or no emission lines, and the majority analyze cluster galaxies at $z\lesssim 1$ \citep{Jorgensen05,Kelson06,SB09,Jorgensen13}, with only few
works addressing the field population \citep{Schiavon06,fritz09,ziegler05,ferreras09}. 

Purely passive evolution of the quiescent galaxy population has been put into question in the work of \cite{Schiavon06} who analyze
the co-added spectra of red-sequence field galaxies at $z=0.8-1$ from the DEEP2 survey \citep{davis03}. From comparison with SDSS stacked
red-sequence galaxy spectra from \cite{eisenstein03} they find that the evolution in the $H\delta$~absorption index is slower than predicted by
passive evolution of a Simple Stellar Population (SSP). The observed evolution could instead be described by a series of continuous star formation models that get
quenched at successive epochs \citep{harker06}. Such evolution is consistent with the modest evolution in the scatter of the color-magnitude relation of quiescent galaxies from
$z\sim1$ to the present-day \citep{ruhland09}. \cite{SB09} analyze cluster and group red-sequence galaxies at $z=0.45-0.75$ in the EDisCS survey \citep{Ediscs} by means
of their co-added spectra in three bins of redshift and two bins of velocity dispersion. They find that the most massive galaxies ($\rm
\sigma_V>175$km/s) are consistent with high formation redshifts and subsequent passive evolution, while lower-mass galaxies require an extended
star formation at low level without inducing metal enrichment. They argue that the observed evolution in index--velocity dispersion relations is
consistent with a scenario in which 40\% of $\sigma_V<175$km/s galaxies enter the red-sequence between $z=0.75$ and $z=0.45$.  Comparison between
field and cluster studies suggests that the mass scale at which simple passive evolution does not apply is different in different environments,
being lower in clusters.
These studies however, based on co-added galaxy spectra, cannot distinguish between an ongoing (low-level) star formation in galaxies that are
on the red sequence already at high/intermediate redshift and the continuous addition at lower redshifts of quenched blue galaxies. 
\cite{Jorgensen13} recently published a very thorough analysis, with good spectral quality for individual galaxies for a total sample of $\sim130$~cluster early-type galaxies at
intermediate redshifts (in particular they study three clusters at $z=0.54, 0.83, 0.89$). The evolution of the Fundamental Plane is consistent
with passive evolution after a mass-dependent formation redshift. However such formation redshifts are rather low ($z_{form}=1.95$ for
$\rm \sigma_V=225km/s$) and inconsistent with the absorption index strengths. Moreover the metallicity and the element abundance ratio of at
least two of the studied clusters differ from that of other clusters both at intermediate and low redshift, implying chemical enrichment in cluster galaxies
and cluster-to-cluster variations.

In this work we aim at quantifying the evolution of the stellar populations of massive galaxies at $z\lesssim1$ in relation to global galaxy
properties such as stellar mass and star formation activity. As an added value over previous works we do not target only quiescent, early-type
galaxies but instead we consider galaxies more massive than $3\times10^{10}M_\odot$ without any pre-selection on their star formation
activity. The goal is to compare the stellar population properties of the massive galaxy population as a whole and of quiescent galaxies alone
to those of the corresponding present-day populations, and explore whether the intermediate-redshift star-forming galaxies can supply the
necessary population for the observed evolution. By targeting a more representative {\it field} galaxy population this work is also
complementary to most previous studies focused on cluster or group galaxies toward a more comprehensive view of galaxy evolution as a function
of environment. 

We select a sample of bright galaxies from the \combo~survey \citep{wolf03,wolf04} with stellar mass $>10^{10}M_\odot$ and redshift $0.65<z<0.75$.
We observe a final sample of 77 galaxies with the Inamori Magellan Areal Camera and Spectrograph (IMACS) on the 6.5 m Magellan telescope at the Las
Campanas Observatory gathering multi-object deep medium-resolution spectroscopy covering the rest-frame optical range from the 4000\AA-break up to
the red Mg and Fe absorption features.  In order to resolve the dispersion in galaxy properties and gain insight into {\it individual} galaxy
evolution (as opposed to evolution of the {\it overall population} through the addition of quenched galaxies),  we work with individual galaxies rather than
co-added spectra. This sample is directly compared with local samples of massive galaxies, extracted from the SDSS with comparable spectral
quality. We perform a consistent analysis of the $z=0.7$ galaxies and of the SDSS samples thus minimizing systematic uncertainties in the inferred
evolution due to different diagnostics and analysis techniques. Specifically, we analyze an optimally-selected set of stellar absorption features
with distinct sensitivity to age and metallicity. Building on our work on SDSS \citep{gallazzi05}, these are interpreted with a large Monte Carlo library of star
formation histories (SFH) applied to the \cite{bc03} simple stellar population (SSP) models and adopting a Bayesian statistical approach in order
to derive full probability density functions of stellar mass, light-weighted mean age and stellar metallicity. 

The sample selection is described in Sec.~\ref{sec:data}, along with the spectroscopic observations and the extraction of redshift, velocity
dispersion, emission and absorption lines. The method to interpret absorption features in terms of physical properties of the galaxy stellar
populations, already
developed and applied to SDSS data, is outlined in Sec.~\ref{sec:physparam_method} and the resulting mass, age and metallicity estimates
are described in Sec.~\ref{sec:physparam_result}. Stellar ages and metallicities are related to galaxy stellar mass and compared to the
corresponding scaling relations for $z=0.1$ SDSS galaxies in Sec.~\ref{sec:results} for the entire population. In
Sec.~\ref{sec:relations_ssfr} we characterize the sample in terms of the galaxy star formation activity and discuss how this affects the
location of galaxies in the stellar population relations and their evolution to $z=0.1$. We discuss our findings in Sec.~\ref{sec:discussion} and
summarize them in
Sec.~\ref{sec:conclusion}. Throughout the paper we adopt a $\Lambda$CDM cosmology with $\rm \Omega_m=0.3$,
$\rm \Omega_\Lambda=0.7$ and $\rm H_0=70~km~s^{-1}~Mpc^{-1}$, and we assume a \cite{chabrier03} IMF.

\section{The Data}\label{sec:data}
\subsection{The sample}\label{sec:sample}
The sample has been selected from the COMBO-17 catalog of the E-CDFS \citep[][ see also Sec.~\ref{sec:combo17}]{wolf04} to have photometric redshift (or spectroscopic when
available) in the range $0.65\leq z\leq 0.75$ and to have stellar mass larger than $\rm 10^{10}M_\odot$ \citep[as estimated from SED fitting
by][]{borch06}. Type-1 QSOs and unobscured AGNs are excluded from our sample based on the spectral classification from the \combo~SED fit.
These criteria give a sample of 717 galaxies (of which 200 already had a spectroscopic redshift from the literature at the time at which the
targets were selected). We further restrict the
selection to galaxies with apparent R-band magnitude $\rm R<22.7$, which allows sampling the red-sequence down to its completeness limit of
$\rm 3\times10^{10}M_\odot$ at this redshift \citep{borch06} and the massive end of the blue-cloud. From the so-selected sample of 521 galaxies
we observed a final sample of 77 galaxies with a single IMACS mask designed in order to give highest priority to galaxies with stellar mass
$\rm M_\star>10^{11}M_\odot$ and then to galaxies with $\rm M_\star>10^{10.5}M_\odot$. The observed targets are listed in
Table~\ref{tab:sample}. 

\begin{figure}
\epsscale{1.2}
\plotone{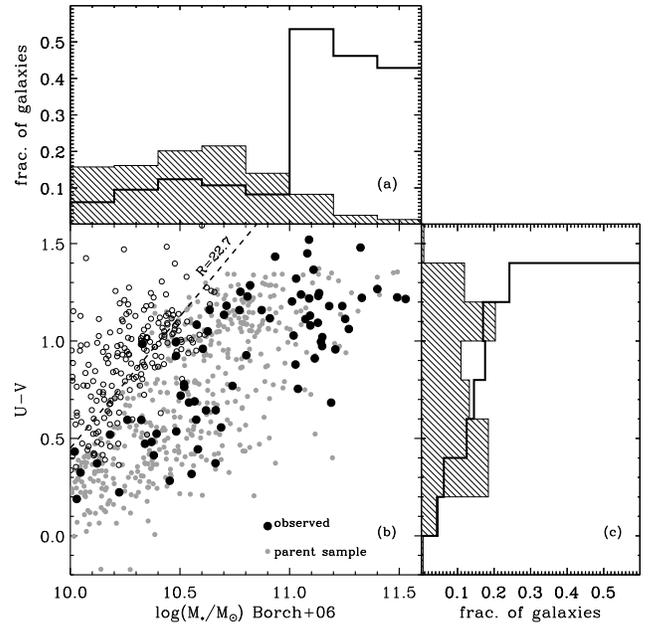}
\caption{\textit{Panel b}: Distribution in rest-frame U-V color versus stellar mass for massive galaxies ($\rm >10^{10} M_\odot$) at redshift $0.65\leq z\leq 0.75$ in
the COMBO-17 catalog of the E-CDFS field (grey dots). Stellar masses are from \cite{borch06} assuming a Kroupa IMF. The sample from which the observed targets have been
drawn is restricted to those galaxies with apparent magnitude $\rm R<22.7$ (empty circles are galaxies fainter than this limit, whose location is shown approximately by
the dashed line). The black filled circles indicate the 77 galaxies observed in our program with IMACS. The distributions in stellar mass (\textit{panel a}) and in rest-frame
$U-V$ color (\textit{panel c}) of the parent redshift- and mass-selected sample (including galaxies fainter than $\rm R=22.7$) are shown with hatched histograms,
normalized to the total number of galaxies. The solid black histograms show instead the fraction of observed galaxies with respect to the number of galaxies in the parent sample
as a function of mass and color.}\label{fig:selection}
\end{figure}

Figure~\ref{fig:selection}b shows the distribution in rest-frame $U-V$ color and stellar mass from \cite{borch06} of the observed sample (black circles)
compared to that of the parent mass- and redshift-selected sample (grey points; empty circles 
are galaxies fainter than $R=22.7$). Figure~\ref{fig:selection}a and c show the histograms in stellar mass and color, respectively, for the parent
sample before the magnitude cut (hatched histograms). The solid black histogram shows instead the corresponding fraction of galaxies observed with respect to the number of galaxies in the parent
sample.
The observed
sample has, by design, a larger fraction of galaxies with masses greater than $10^{11}M_\odot$ and a lower fraction at masses below
$10^{10.5}M_\odot$ compared to the parent sample. We have a
completeness between 40 and 50\% at masses $>10^{11}M_\odot$ which drops around 10\% at masses between $10^{10.5}$ and $10^{11}M_\odot$, and below 10\%
at lower masses. 
The distribution in color reflects the bimodality of the parent sample, but with a
larger fraction of red-sequence galaxies, induced by the priority given to high-mass galaxies. 

This figure illustrates how representative the observed spectroscopic targets are with respect to the underlying parent distribution of
galaxies, based on photometric information provided in \cite{wolf04} and \cite{borch06}. 
The figure is basically unchanged if we use the updated photometry from \cite{wolf08} and stellar masses based on \citet{bell03} formula, except
that the masses are $\sim0.1$dex larger due to the updated photometric calibration. We also note that the statement that our sample is strictly
above $10^{10}M_\odot$ and more representative of galaxies more massive than $10^{10.5}M_\odot$ is preserved with our stellar masses derived from
accurate spectroscopic redshift, spectroscopic diagnostics and more complex modeling (as shown in Sec.~\ref{sec:physparam_result}).

\subsection{The observations}\label{sec:observations}
We obtained deep, medium-resolution rest-frame optical spectroscopy of our sample galaxies with IMACS on
Magellan 6.5m Baade Telescope. We used the f/2 camera which covers a field of view of $\rm 27.5\times27.5~
 arcmin^2$, the red 300 l/mm grism with the filter WB6300-9500 and with slits 1 arcsec wide and 15 arcsec long,
which provide a spectral resolution of 6.25\AA~FWHM ($\sim 3.7$\AA~rest-frame) constant over the wavelength range
6500--9500\AA. This covers the rest-frame range between $\sim3700$\AA~and $\sim5500$\AA~where the most
important absorption features for stellar population analysis are located, including the 4000\AA-break up to
Fe5335. The spectral resolution is comparable to that of SDSS spectra and to the \cite{bc03} models that we
adopt  for the analysis.

We observed a final sample of 77 galaxies which is the number of slits that could be placed in a single IMACS mask given our resolution and
slit length. In addition to these, we simultaneously observed in the same mask three bright blue stars for telluric correction, which were
selected from the COMBO-17 catalog to have a `Star' SED class, R-band magnitude $\rm 17<m_R<18$ and colors $B-V<0.2$, $V-R<0.1$. Observations of a photometric
standard star (Feige 110) were taken at the beginning and at the end of  each night. 

The objects were observed for a total exposure time of 10 hours, split into several exposures of at most 30 minutes
each. Dithering of a few arcsec along the slit was applied at each exposure, in order to improve sky subtraction and bad
pixel removal. Spectroscopic flats and arcs were taken in between sequences of three science exposures.  The
observations were carried out over three consecutive nights (17-18-19/10/2009), with typically clear sky conditions
and average seeing of 1" for the first two nights and 0.6" for the third one.

Data reduction has been performed using version 2.14 of the {\sc cosmos} package, a set of programs optimized for the reduction of multi-slit spectra obtained with IMACS on the
Magellan Telescope. The reduction process consists of the alignment of the slit-mask relative to the CCD focal plane and the construction of an accurate spectral map using arc
spectra that bracket each set of three (dithered) science exposures. Standard reduction steps are followed to subtract bias and divide by spectroscopic flats. Optimal sky
subtraction is achieved in two steps: first, the background is subtracted from all the 2D spectra in a science exposure using the {\sc subsky} routine in {\sc cosmos}, which
uses the \cite{subsky} procedure to construct a cosmic ray-cleaned median sky spectrum for each slit relying on the CCD-science coordinate system mapping; then individual 2D
spectra are extracted; in the second step, pairs of consecutive dithered 2D spectra are subtracted from each other to remove residuals\footnote{For four objects (obs074,obs029,
obs051, obs066) dithered-subtracted frames  could not be used for the presence of a nearby object whose trace would affect the target trace in the difference 2D spectra.}.  We
then use the standard IRAF routine {\sc apextract} to extract 1D spectra from the background subtracted 2D spectra while refining the background subtraction using two `sky'
regions adjacent to the continuum trace. Individual spectra in each science exposure may be too faint to robustly trace the galaxy continuum. As the mask is designed so to put
all galaxies at the same relative position with respect to their own slit, we can safely define the continuum  trace on a combined spectrum from all the slits in the mask and
use this common trace to extract the individual spectra.

The extracted 1D spectra are corrected for telluric
absorption in the following way: we derive a smooth fit of the 1D spectra of the three telluric standard stars
in the spectral mask and calculate a transmissivity function by dividing the observed spectrum by the smooth
fit, excluding regions of intrinsic stellar absorptions. We take the median of the functions obtained from the three
standard stars and finally divide all the 1D spectra in the same exposure by this transmissivity function. 
Similar reduction steps are followed for the spectro-photometric standard star (Feige 110) observations and a
sensitivity function is obtained with the {\sc standard} and {\sc sensfunc} IRAF routines. Each 1D spectrum is
spectro-photometrically calibrated applying the sensitivity function of the closest in time standard star observation.

The flux calibrated spectra of each galaxy are finally combined with an exposure-time-weighted median. The associated error spectrum is
computed as half of the $16^{th}-84^{th}$ inter-percentile range of the exposure-time-weighted flux distribution, divided by the square
root of the number (25) of frames combined. The average S/N per resolution element (corresponding to $\sim 3.7$\AA~rest-frame), over the entire wavelength range
3800--5500\AA, for the whole sample is
$\sim 19$ (or 11 per pixel), with 43\% (85\%) of the spectra having $S/N>20$ (10) per resolution element.

In the analysis we exclude four objects (obs173, obs443, obs444, obs515) because their spectrum was either severely contaminated by a nearby 
object or of very poor quality hampering any index measurement. The final analyzed sample is thus composed of 73 objects. Figure~\ref{fig:spectra} shows as example the 
spectra of four galaxies with different star formation properties and at different spectral S/N ratio.

\subsection{Data processing}\label{processing}
\subsubsection{Redshift}
We measure the redshift using the {\sc FXCOR} procedure in IRAF by cross-correlating the galaxy stellar continuum with a
set of template spectra composed of SSPs of different ages and metallicities. The typical redshift uncertainty
is $10^{-4}$ and not greater than 0.0004. Photometric redshifts from COMBO-17 differ on average by $-0.015$
from the spectroscopically derived redshifts with a scatter of 0.047. For the subset of 34 galaxies for which a
spectroscopic redshift measure was already available in the literature we find that our redshift determination
agrees with those with a scatter of 0.002. According to our spectroscopic redshift measurements, the observed 
sample spans the redshift range $0.65<z<0.77$ with an average redshift of $0.696$ and a standard deviation of 0.03.

\subsubsection{Stellar velocity dispersion}\label{vdisp}
A reliable measurement of the stellar velocity dispersion is important for a proper comparison of the population synthesis models 
with the observed spectra and absorption line diagnostics to infer the stellar population physical parameters. 
The spectral resolution of our data is 6.25\AA~FWHM observed-frame, corresponding to $R=1280$ at the blaze wavelength of 8000\AA, which allows us to determine stellar
velocity dispersions down to $\approx100$ km/s. We use the publicly available pPXF code \citep{ppxf} to fit the stellar continuum with a set of
template spectra based on \cite{bc03} SSP models and solve for the kinematics (systemic velocity and velocity dispersion). The models are first
convolved to match the spectral resolution of the data, and both models and data are rebinned on the same logarithmic scale. We repeat the fit
100 times on Monte Carlo realizations of the spectrum perturbed according to its noise. We take the mean of the results as our fiducial
estimate of systemic velocity and velocity dispersion, and the standard deviation as uncertainty on these parameters. The quality of the data
allows us to derive measurements of velocity dispersion for 84\% of the sample (61/73 galaxies). For the remaining objects the uncertainties on
velocity dispersion are larger than 100 km/s or larger than the value itself (these galaxies have masses lower than $\sim10^{11}M_\odot$ and their spectra have $S/N<15$ per
resolution element).
The typical accuracy on the velocity dispersion $\sigma_V$ is 19\%,
improving to 6\% for spectra with $S/N>20$ per resolution element. For the subset
of galaxies for which we can fit the kinematics the average velocity dispersion is 184~km/s with a standard deviation of 71~km/s. We have
tested that we find consistent results and uncertainties using a different template set based  on the MILES stellar library
\citep{miles06,miles11} instead of BC03 SSPs.

\subsubsection{Emission lines}\label{continuum_fit}
\begin{figure*}
\epsscale{0.8}
\plotone{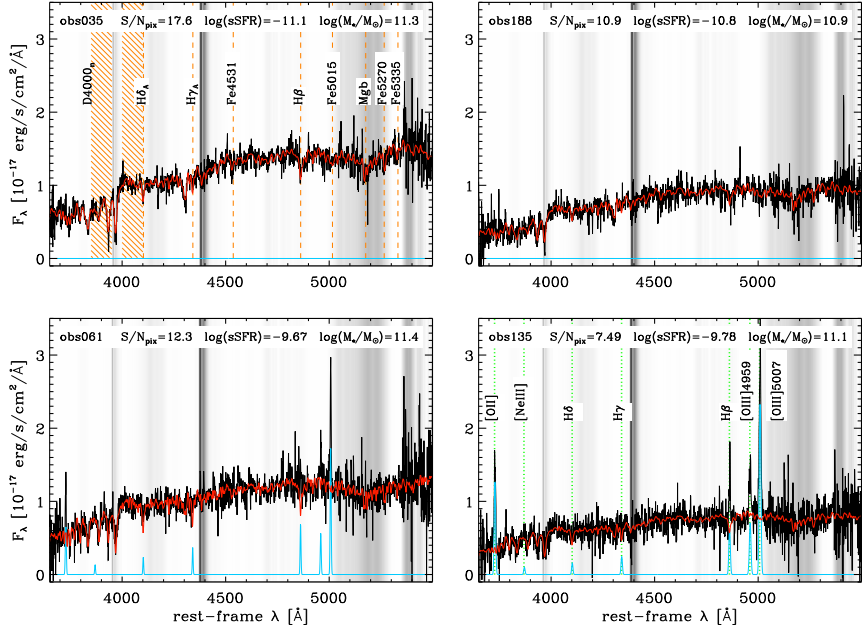}
\caption{Observed spectra of four galaxies in the sample, two quiescent galaxies (upper panels) and two star-forming galaxies (bottom panels) at 
different spectral S/N per pixel (1 pixel $\sim 1.2$\AA). The red spectrum shows the best-fit model of the stellar continuum as obtained with {\sc platefit}, while the cyan line indicates 
the emission lines detected at more than $3\sigma$. The central wavelength of some emission lines and of the absorption features used in the analysis is marked 
with dotted green lines (bottom-right panel) and with dashed orange lines (upper left panel) respectively. In each panel, the grey intensity is proportional to the 
average atmospheric transmissivity function, whereby white is 100\% transmission while darker areas are the wavelengths more affected by telluric absorption.}\label{fig:spectra}
\end{figure*}

The derivation of physical parameter estimates of galaxy stellar populations relies on the analysis of the
stellar continuum and stellar absorption features in the optical galaxy spectra. It is therefore necessary to
first separate the stellar continuum from any nebular emission lines. To this purpose we make use of the
{\sc platefit} code which was developed in \cite{Tremonti04} and \cite{Jarle04} to analyze the emission line spectra
of SDSS galaxies. The code performs a non-negative least-squares fit to the emission-line-free regions of the
spectrum finding the best fitting linear combination of SSPs and the best fitting attenuation A$_V$. The SSPs
are based on the \cite{bc03} models and have different metallicities ($Z_\star/Z_\odot$ = 0.2, 0.4, 1, 2.5) and
ages (between 5.2~Myr and 7.16~Gyr, i.e the age of the Universe at $z=0.7$). Before fitting, the observed
spectra are corrected for foreground reddening using the \cite{Schlegel98} maps and put in rest-frame, while the model spectra are broadened to
match the data spectral resolution and the galaxy stellar velocity dispersion. The best fit {\it continuum} is
then subtracted from the original spectrum and any remaining residuals smoothed over a 200-pixels scale
({\it smooth\_resid}) are
further removed. The final residuals are fitted with Gaussian-broadened emission line templates. We allow the
broadening width of the lines to vary within $\pm50$~km/s of the measured stellar velocity dispersion (or over the larger
range 50-500 km/s for galaxies without a reliable estimate of $\sigma_V$). Emission lines with at least a
3-$\sigma$ detection ({\it nebular} spectrum) are removed from the original spectrum in order to obtain a `pure' stellar continuum
spectrum. Examples of the continuum and emission lines fits are illustrated in Fig.~\ref{fig:spectra}.

Only 20 galaxies have spectral coverage for the \oii~ emission line and among these 10 have a detection above
3$\sigma$. The \oiii5007 line is detected in 26 galaxies, while the \oiii4959 in 16 galaxies. The Balmer
emission lines are detected above 3$\sigma$ in 40, 20, 19 galaxies for \hbeta, H$\delta$ and H$\gamma$
respectively.

\begin{figure*}
\epsscale{0.8}
\plotone{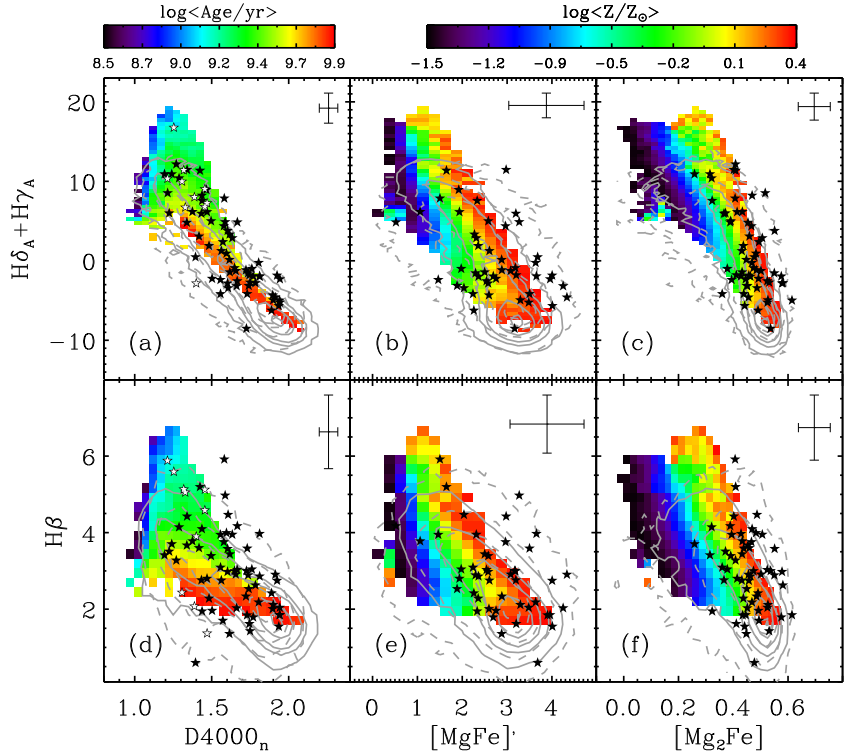}
\caption{Distributions in the five absorption indices used in the analysis. In each panel the starred symbols show
the galaxies in our $z=0.7$ sample for which both indices can be measured from the emission-line corrected
spectrum. Galaxies that lack both Mg-Fe composite indices are shown as white stars in panels a and d. 
The solid grey density contours show the distribution of SDSS galaxies in the redshift range
$0.05<z<0.22$ and with spectral S/N$>20$. 
The dashed contours refer to a subsample of SDSS galaxies that matches the stellar mass distribution of our $z=0.7$ sample, with 
age younger than the Universe age at z=0.7 and no cut in spectral S/N. The contour levels enclose 90, 60, 30, 10, 1\% of each subsample.
The error bars in each panel display the average observational error on
the indices for our $z=0.7$ sample. The underlying color maps show the distribution of all the models in our SFH library
with $\rm t_{form}$ younger than the age of the Universe at $z=0.7$. The color coding reflects the median
light-weighted average age in each index-index bin in the panels showing the Balmer absorption features against
\dn~ (panels a and d) and the median stellar metallicity in the panels showing the Balmer absorption features
against the composite metal-sensitive indices (panels b,c,e,f).}\label{fig:idx_plane}
\end{figure*}

\subsubsection{Stellar absorption features}\label{continuum_fit2}
We measure stellar absorption features on the emission-line corrected spectra adopting the band-pass definitions of the Lick system
\citep{worthey94}, including the narrow definition of the 4000\AA-break, \dn~ \citep{balogh99}, and the high-order Balmer lines \hda, \hga~
\citep{wo97}. The wavelength coverage of the IMACS spectra allows us to measure absorption line indices up to Fe5335 for all the galaxies in the
sample (see upper-left panel of Fig.~\ref{fig:spectra}). In the following analysis we are particularly interested in the age-sensitive indices \dn, \hbeta, \hdg~and the metal-sensitive indices
\mgtwofe~and \mgfep.\footnote{The composite indices are defined as \mgfep $= \rm \sqrt{Mgb~(0.72~Fe5270+0.28~Fe5335)}$ following \cite{thomas03},
and \mgtwofe $= \rm 0.6~Mg_2 + 0.4~log(Fe4531+Fe5015)$ following \cite{bc03}.}  The combination of Mg and Fe features in these composite indices is designed to remove the
dependence on [$\alpha$/Fe] abundance ratio which is assumed to be solar in the \cite{bc03} models used here. The observational error on the indices accounts for both the noise
spectrum and continuum shape uncertainties associated with either the spectro-photometric calibration (which can affect in particular the
4000\AA-break) or the telluric correction (which can affect in particular the reddest Fe indices). 
We estimate the bias on the index values induced by the continuum shape uncertainty as the difference between the indices measured on the
best-fit {\it continuum} plus the large-scale residuals {\it smooth\_resid} and those measured on the best fit {\it continuum} only (the unbiased
estimate). We correct the index measured on the observed spectrum by subtracting this bias and we include the absolute value of the bias in the
error budget of the index.
This correction is negligible for \hb, while
it is typically 15\%, 10\% and 25\% of the index formal error for \hdg, \mgfep~and \mgtwofe~respectively. The correction is significant for
\dn~for which it is on average 1.4 times the index formal error. This is in fact expected, since the error budget for \dn~is dominated by flux
calibration uncertainties on scales of the order of the two band widths, rather than by uncorrelated noise in individual pixels, as is the case for
the other indices, which have a narrower wavelength span.

The distribution of our sample in index-index diagnostic diagrams is shown in Fig.~\ref{fig:idx_plane} (black stars).
Panels (a) and (d) show the Balmer absorption features against \dn~ as diagnostics of the mean stellar age and recent
star formation history. The color maps indicate the distribution of model spectra constructed assuming a large range of SFHs (see Sec.~\ref{sec:physparam_method}) 
and the color coding reflects their light-weighted mean age. 
The other panels show instead combinations of Balmer absorption features and metal-sensitive
indices which alleviate the degeneracy between mean age and stellar metallicity. The color maps in these panels illustrate the mean stellar metallicity of 
the models in our library. In each panel we show only the
galaxies in the sample for which the two corresponding indices can be measured off the emission line corrected
spectrum. The error bars indicate the associated average uncertainty on the index strengths. For comparison we
show also the distribution of SDSS DR4 galaxies at redshift $0.05<z<0.22$ with median S/N per pixel greater than 20
(grey contours). We note that our sample galaxies do not reach the high \dn~and low Balmer absorption end of the
SDSS local sample, reflecting the younger ages of the $z=0.7$ galaxies. They also show a larger spread due to the lower 
average S/N, as suggested by a mass-matched SDSS subsample with stellar age younger than the Universe age at $z=0.7$ and no 
cut in S/N (dashed contours).
Our sample also tends to
be distributed at the highest values of the metal-sensitive indices. One reason for the lack of low values of the \mgfep~and \mgtwofe~indices 
in our sample can be the fact that these composite 
indices are more likely mathematically ill-defined in the low-S/N spectra of galaxies with intrinsically weak Mg and
Fe absorptions that would scatter to negative values: indeed the galaxies for which we cannot measure neither \mgfep~nor \mgtwofe~are young 
galaxies (white stars in panels a and d).

\subsection{Ancillary data}\label{sec:multilambda}

The observed sample has several ancillary multi-wavelength data, ranging from the UV/optical from \combo~(the catalog used for the selection
of the sample), to the NIR from HAWK-I and to the FIR from Spitzer. In addition, one-orbit depth HST optical imaging in F606W and F850LP is
available for 85\% of the sample from the GEMS survey \citep{gems}. Here we briefly describe the ancillary data that will be used in the analysis.

\subsubsection{COMBO-17}\label{sec:combo17}
\combo~has surveyed an area of
$\sim 34' \times 33'$ in the Chandra Deep Field South
to deep limits in 5 broad and 12 medium passbands \citep{wolf03,wolf04,wolf08}.
Objects are classified fitting 
non-evolving galaxy, star, and AGN template spectra to the SEDs. Photometric redshifts are assigned for $\sim 99$\% of the
objects to a limit of $m_R \sim 23.5$ with a typical accuracy
of $\delta z/(1+z) \sim 0.02$ \citep{wolf04}. This 
allows construction of $\sim 0.1$ mag accurate
rest-frame colors and absolute magnitudes \citep[accounting for distance
and $k$-correction uncertainties; see also][]{wolf08}. Astrometric accuracy is $\sim 0.1$\arcsec. 

As described in Sec.~\ref{sec:sample}, SED classification, photometric redshifts, absolute R-band magnitude from \combo~have been used to preselect
the target sample. In addition, we will use the luminosity in the \combo~synthetic band centered at 2800\AA~to estimate the amount of (unobscured)
star formation in the observed galaxies.

\subsubsection{HAWK-I NIR imaging}\label{sec:hawki}
Deep and high-resolution NIR imaging in $K_\mathrm{S}$ and $J$ band
was obtained with HAWK-I
\citep{2004SPIE.5492.1763P,2008A&A...491..941K,2006SPIE.6269E..29C} at
the ESO-VLT in program 082.A-0890.  The full E-CDFS field was covered
with a $4\times 4$ tile mosaic of dithered exposures for a total of 65
min on-target exposures in $K_\mathrm{S}$ and 45 minutes in $J$. Sky
conditions were nearly excellent for most of the exposures, yielding
PSF FWHM$\lesssim 0.5$~arcsec in $K_\mathrm{S}$ and $\lesssim
0.8$~arcsec in $J$ over the entire mosaic. Data reduction has been
performed by S. Z. using a customized pipeline based on the original
version distributed by ESO, which implements improved recipes for the
construction of the master flat field and for the frame coaddition to
properly take into account object masks and variance maps. The
implementation of object masks is particularly relevant for the
photometry of this work, since it eliminates the effects of background
over-subtraction which is often seen in the neighborhood of (bright)
sources.  Before the final mosaic, the individual tiles of stacked
exposures have been photometrically calibrated with a two-step
strategy. A short mosaic of the field was observed in photometric
conditions immediately before or after a standard star observation;
this shallow calibrated mosaic was then used as reference for
secondary calibration of the full-depth tiles, based on the brightest
unsaturated stars. Typical field-to-field zero point fluctuations are
$\lesssim 0.05$~mag. We estimate a 5-$\sigma$ point source limiting
magnitude of $K_\mathrm{S}$(AB)$\sim24.3$ ($\sim 24.4$ in $J$).

Fluxes for the targets of this study were obtained from SExtractor's
\citep[version 2.8.6,][]{sextractor} AUTO\_FLUXes, adopting Kron\_fact=3.0
and min\_radius=3.5 for both $J$ and $K_\mathrm{S}$: we have checked
that these parameters produce integrated fluxes as close as possible
to ``total'', while yielding an optimal SNR.  

In the following analysis we will use the $J$ and $K_\mathrm{S}$ HAWK-I 
data, in combination with spectral information, to put constraints on dust and stellar masses. 

\subsubsection{FIDEL Spitzer 24\micron~ data}
In order to estimate the total star formation rate (including the dust-obscured contribution) in our sample, we incorporate deep Spitzer 24{\micron}
imaging from the FIDEL survey of the Extended Chandra Deep Field South \citep{dickinson07,magnelli09}. At the 6{\arcsec}
resolution of Spitzer/MIPS 24{\micron} observations, galaxies at $z \sim 0.7$ are unresolved, and point source
detection is performed using STARFINDER on the FIDEL 24{\micron} dataset; aperture photometry is performed on the image
when nearby sources have been PSF-subtracted away. The 3 (5) $\sigma$ detection limit is 24.6 (42) $\mu$\,Jy, corresponding
to the 50\% (80\%)  completeness limit (X.-Z.\ Zheng et al., in preparation). 
After matching the FIDEL catalog down to the 3$\sigma$ detection limit to the \combo~coordinates of our sample with a 1" matching radius, these 24{\micron} fluxes are combined
with the COMBO-17 rest-frame near-UV photometry to estimate a total (obscured and unobscured) star formation rate,
following \cite{bell05} and \cite{gallazzi09}.

\section{Physical parameter estimates}\label{sec:physparam}
\subsection{Method}\label{sec:physparam_method}
In order to estimate stellar population physical parameters, such as stellar mass, light-weighted mean age and
stellar metallicity, we adopt a Bayesian approach as developed and applied to the analysis of SDSS galaxies in
\cite{gallazzi05}, in which the observational data are compared to a large library of model spectra generated
from prior assumptions on the possible galaxy SFHs. We wish to compare the physical parameters of
$z=0.7$ galaxies with those of local SDSS galaxies and infer their evolution over this time span. 
Therefore we decide to perform an analysis as close as possible to the one performed on SDSS galaxies in order to minimize
systematic differences. 

Without an appropriate knowledge of how the true SFH distribution varies over this redshift interval, we  take the more conservative approach
of adopting the same Monte Carlo model library constructed in \cite{gallazzi05} based on \cite{bc03} SPS models. We address in 
Appendix~\ref{sec:physparam_systematics_prior} the possible effects of changing the SFH prior. We anticipate that our conclusions remain valid even
with a different SFH prior, but with small quantitative differences. In the adopted model library, the SFHs are modeled with exponentially
declining laws on top of which random bursts can occur at any time with a probability such that 10\% of all the models experience a burst in the
last 2 Gyr. They are parameterized in terms of the fraction A of mass produced in the burst with respect to the total mass produced by the
exponential model, at a rate that is assumed constant for the duration of the burst. The metallicity of the models can vary logarithmically in the range
$0.02-2.5\times Z_\odot$ but is kept fixed along each SFH. Because the strength of several absorption features depends on the broadening due to
stellar velocity dispersion (in addition to the spectral resolution) the model spectra are also convolved with a range of velocity dispersions in
order to match the effective broadening of the observed spectra. The range of model parameters is summarized in Table~\ref{tab:library}.
  
The observational constraints are given by a set of five absorption indices, chosen to have distinct sensitivities to age (\dn, \hbeta, \hdg) and
metallicity (\mgfep, \mgtwofe) and minimal dependence on element abundance ratios. For each galaxy the observed index strengths are compared to
those predicted by all the models in the library with a velocity dispersion consistent within the errors with the galaxy velocity dispersion
\footnote{For
galaxies for which a reliable estimate of $\sigma_V$ cannot be derived we consider the full range of velocity dispersions in the library. In
practice this has negligible effect on the derived parameter estimates, since at the S/N of these galaxies ($<10$) the differences in index strengths due to velocity
dispersion differences are well within the index uncertainties.}. We then
construct the probability density function of the parameters of interest by weighing each model by its likelihood $\rm exp(-\chi^2/2)$ and
marginalizing over all the other parameters. For 43/73 galaxies all the five selected indices are measured. For the other galaxies we restrict the
fit to the subset of indices measured: for 16 galaxies we can use all the age-sensitive indices and either \mgtwofe~ or \mgfep; for one object we
lack only \hdg; for 3 galaxies we can use \dn~ and \hdg~ in combination with one of the metal-sensitive indices. We checked that none of these
cases affects our results (see Appendix~\ref{sec:physparam_systematics_indx}). Finally, for 10 galaxies we
can only use the age-sensitive indices. As discussed in Appendix~\ref{sec:physparam_systematics_indx} the metallicity estimates of these galaxies is
very uncertain and may be biased low. Therefore they will be flagged in the following plots and excluded by default from the analysis, unless otherwise
stated. We note though that we do not find evidence for a bias in their light-weighted ages.

Figure~\ref{fig:idx_plane} illustrates the coverage of the model library in the indices analyzed and the predicted 
variations in mean light-weighted stellar age along the \dn-Balmer lines planes (panels (a) and (d)) and in mean
stellar metallicity in the planes combining Balmer lines with composite Mg-Fe indices (panels (b), (c), (e), (f)). The figure shows that the
models cover well the distribution of the data in these diagnostic diagrams and it indicates that our $z=0.7$ sample
span a relatively large range of ages, comparable to the distribution of low-redshift SDSS galaxies (except for ages
older than the age of the Universe at $z=0.7$, as expected), and a relatively narrow range in metallicities, in
particular for galaxies with strong Balmer absorptions.

\subsubsection{Stellar mass and dust attenuation}
Stellar masses are obtained by normalizing the intrinsic mass-to-light ratio of each model to the observed,
dust-corrected K-band luminosity from HAWK-I (which roughly corresponds to the rest-frame J). Attenuation by dust in
the observer-frame K-band is estimated from the difference between the observed $J-K$ galaxy color and the color measured on
the dust-free, redshifted model spectra and adopting a single power-law attenuation curve \citep[$\propto$
$\lambda^{-0.7}$,][]{CF00}\footnote{\cite{Chevallard13} argue for a more complex attenuation curve in which the slope is a function of the
V-band optical depth to account for the effect of
inclination and spatial distribution of dust in unresolved galaxies. We have estimated that by adopting their prescription would have a
small impact on the inferred stellar masses for galaxies with $A_{K,obs}\lesssim0.5$. The stellar mass could vary between $-0.1$~dex and
0.1~dex in more dusty galaxies.}. In doing so, for each galaxy we exclude those models that would predict unphysically negative dust
attenuations. However, imposing a strict cut at $A_{K,obs}>0$ would not account for the observational errors on the color and model uncertainties 
in the absorption indices-color correspondence (see also sec. 2.4.3 of Gallazzi et al. 2005). We thus include models producing a dust attenuation 
down to $A_{K,obs} \gtrsim -0.1$ mag.
This approach is
similar to the one adopted for the analysis of SDSS galaxies, in which the fiber $r-i$ color (corrected for emission
lines) was used to estimate dust attenuation \citep{kauffmann03,gallazzi05}. For the analysis of our $z=0.7$ galaxies
we lack the spectral coverage to correct for emission lines colors corresponding to the rest-frame $r-i$,
therefore we use an optical-NIR color redward of H$\alpha$ ($J-K$ roughly corresponds to rest-frame $I-J$) where no
emission line is expected to significantly affect the broad-band magnitudes. We note that the condition $A_{K,obs} > -0.1$~mag 
adopted here corresponds to slightly more negative attenuations than allowed in \cite{gallazzi05} ($A_z>-0.1$~mag) but we have 
checked that this has a negligible impact on the derived parameter estimates.

The IMACS 1" slit samples on average 70\% of the total I-band flux from \combo~(the fraction varies between 40\% and 100\%), while the $J-K$ color
used to estimate dust attenuation is representative of the whole galaxy. This may introduce some mismatch between the stellar population properties
and the dust attenuation if significant color gradients are present beyond the effective radius. A precise quantification of this effect
requires a careful convolution of the images with the actual effective PSF in the spectra. However simple considerations exclude that this is a
major effect in our analysis. First of all, a posteriori we do not detect any trend between the inferred dust attenuation and the fraction of light
sampled by the spectrum as would be the case if there were systematic color gradients. Secondly, our spectra are not only sensitive to a central
aperture but include the full extent of the galaxy along the slit direction: this attenuates the possible aperture bias. Thirdly, assuming that
$\sim70$\% of the total K-band flux is in the slit and assuming a difference in color of 0.5 (0.2) mag between the regions inside and outside the
slit, this would result in 0.13 (0.06) mag difference between the total $J-K$ color and the one observed in the slit. Considering that 0.5 mag is the
dynamic range in the observed $J-K$, we can exclude that our colors are biased by more than 0.13 mag.

We also give an estimate of the potential different aperture bias affecting our $z=0.7$ data and SDSS in
Appendix~\ref{sec:physparam_systematics_aperture} and we note that the estimated potential bias would not change our conclusions.

\begin{figure*}
\epsscale{1}
\plotone{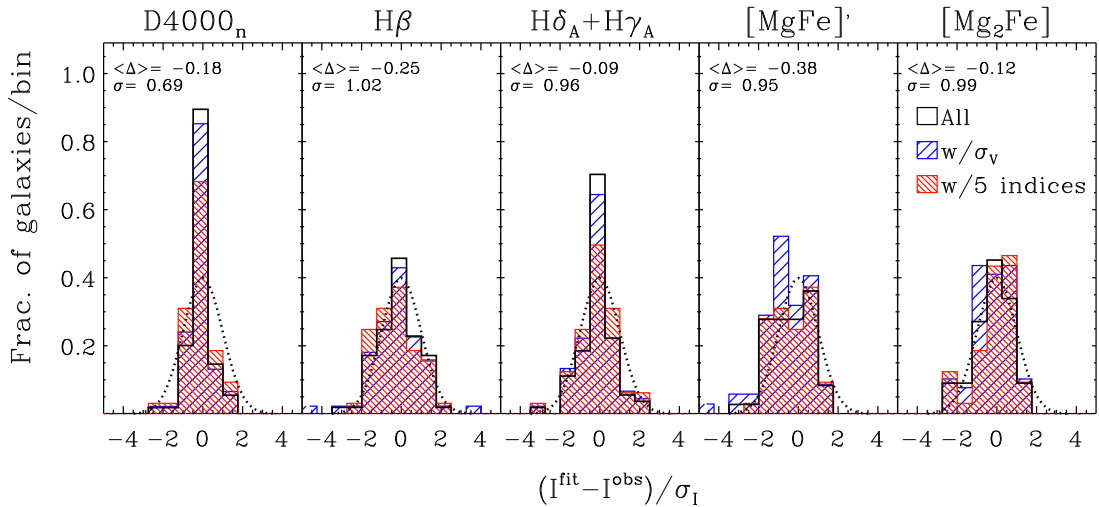} 
\caption{Difference between the absorption index strengths of the best fit model and those measured on the observed spectrum,
normalized by the observational error, for each of the five indices used in the fit. In each panel only those galaxies for which the 
corresponding index can be measured are shown (black empty histograms). The blue hatched
histograms represent the subsample of galaxies with a measure of velocity dispersion, while the red hatched histograms represent
the subsample with a measure of all the 5 indices analyzed. The labels in each panel provide the mean normalized difference and
the standard deviation of the distribution for galaxies with all the 5 indices. A Gaussian centered on zero and with unit standard deviation 
is shown for reference (dotted curve).}\label{fig:didx}
\end{figure*}
\subsubsection{"Goodness of fit" assessment}
Figure~\ref{fig:didx} illustrates how well our models can reproduce the observed index strengths. Each histogram shows the distribution
of the difference between the index strength of the best fit model and the observed index, normalized to the observational error
(black empty histograms). The blue histograms refer to galaxies for which we can measure the velocity dispersion (61/73), while the red histograms
refer to galaxies with all the five indices available (43/73). The distributions for all the indices are in good agreement with the expected
Gaussian distribution centered on zero and with unit standard deviation (hence consistent with the observational error). The quality of the fit is comparable to
SDSS \citep[see Fig. 2 in][]{gallazzi05}, with a slight tendency
towards negative offsets\footnote{This tendency is reproduced, at least for
\dn, \mgfep~and \mgtwofe, in the subsample of SDSS galaxies that matches the mass distribution of our sample, with age younger than the Universe at $z=0.7$ and 
extending to $\rm S/N<20$.}. This is driven by those galaxies outside of (but still consistent within the errors with) the model grid in Fig.~\ref{fig:idx_plane}. We checked that
the derived scaling relations discussed later are not significantly affected by these galaxies. We also obtain very similar distributions when we use the median of the PDF of
each index instead of the best fit model. We have tested that the specific assumptions we make on the prior do not affect
significantly the recovery of the index strengths. In particular, we find statistically indistinguishable distributions if we allow for formation age older
than the age of the Universe, or we modify the model library assuming a larger fraction of recent bursts and extending the star formation
timescales down to 0.1~Gyr (see also Appendix~\ref{sec:physparam_systematics_prior}).

\subsection{Results}\label{sec:physparam_result}
\begin{figure*}
\epsscale{0.8}
\plotone{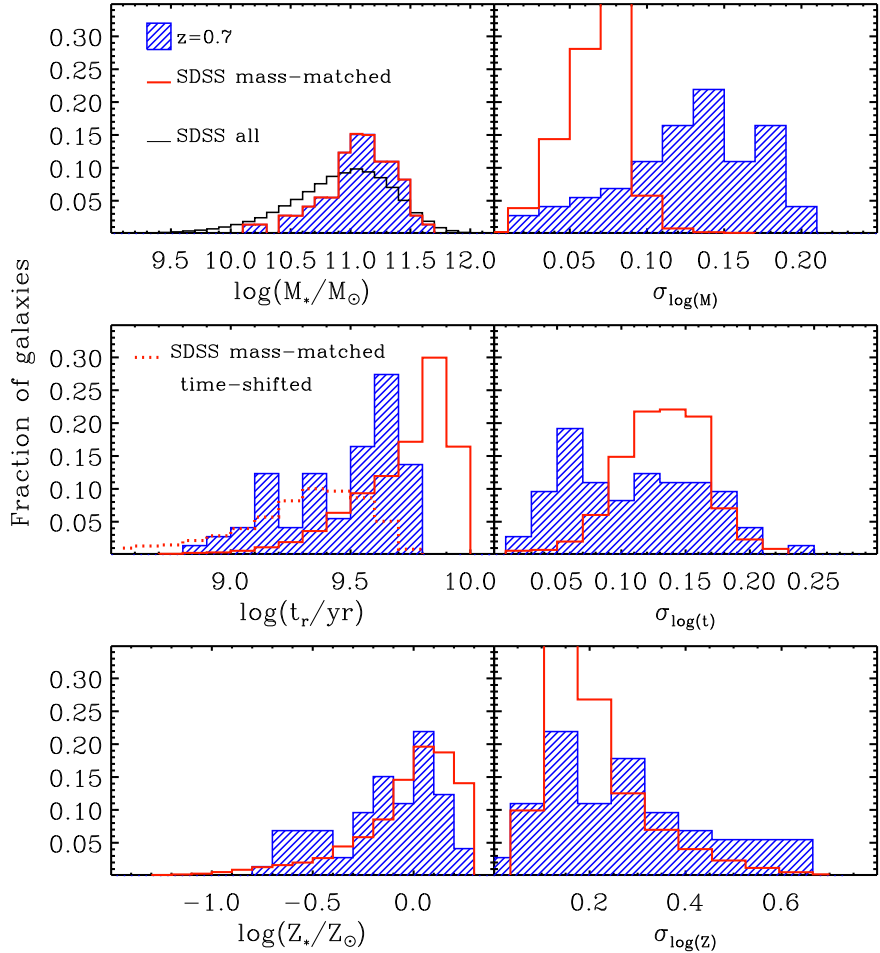}
\caption{Distribution in stellar mass, r-band light-weighted mean age and stellar metallicity (left-hand panels, from top to bottom), estimated
from the median of the corresponding PDFs, and in their associated uncertainties, estimated from the $84^{th}$ and $16^{th}$ percentiles of the
PDFs (right-hand panels). The hatched blue histograms show the distributions of the physical parameters estimated for the $z=0.7$ galaxies, in
comparison to the corresponding distributions for a subsample of low-redshift ($0.05<z<0.22$) SDSS galaxies randomly drawn in order to match the
stellar mass distribution of our $z=0.7$ sample (red histograms; the black histogram in the upper-left panel shows the mass distribution of the
whole SDSS sample). In the middle left panel the
dotted red histogram shows the distribution of the mass-matched SDSS subsample obtained after shifting their ages by the look-back time since $z=0.7$. All
histograms are normalized to unit integral.}\label{fig:param_distr}
\end{figure*}

The derived constraints on the physical parameters for the $z=0.7$ sample are shown in Fig.~\ref{fig:param_distr} (blue hatched histograms). The
average stellar mass of our sample is $\log(M_\star/M_\odot)=11.04$; the distribution in light-weighted mean ages shows a peak at $\sim4.6$~Gyr and
a significant tail down to $1-2$~Gyr. Our sample shows a quite broad distribution in stellar metallicity, with a
peak around solar and a declining tail down to metallicities of $0.1 Z_\odot$. It has a
similar average stellar mass and a similar distribution in stellar mass at $\log(M_\star/M_\odot)\geq11$ as the $0.05<z<0.22$ SDSS sample (black
histogram in the upper-left panel), but the fraction of galaxies with lower
masses drops more quickly in our sample compared to SDSS. For comparison purposes, we consider a subsample of SDSS galaxies randomly drawn to match
the mass distribution of our IMACS sample (red histograms). As expected the SDSS sample peaks at older ages ($\sim$7~Gyr).
By shifting the ages of the SDSS mass-matched sample by the time elapsed between $z=0.7$ and $z=0.1$ (the average SDSS redshift) we notice
interestingly that the resulting distribution (red dotted histogram) has a much lower fraction of old galaxies than the observed $z=0.7$
distribution. We discuss this issue in more depth in Sec.~\ref{sec:relations_all_age}. Finally, we notice that the local SDSS mass-matched sample has a similar metallicity
distribution as the $z=0.7$ one.

The right-hand panels of Fig.~\ref{fig:param_distr} show the distribution in the uncertainties on the derived
parameters. Stellar ages are constrained within 0.11~dex on average and typically better than 0.2~dex, similar
to SDSS galaxies. The average uncertainty on stellar metallicity is 0.3~dex, slightly worse than for SDSS. The uncertainties on stellar mass are typically 0.13~dex,
about twice the average uncertainty on stellar masses for low-redshift galaxies. This is due to the fact that the stellar M/L
varies faster at younger ages, so the same error on age translates into a larger error on M/L. Moreover, the larger error on
stellar mass is also due to the large uncertainties on dust attenuation (the average error on $A_{K,obs}$ is 0.24~mag, compared to the average
error of 0.09~mag on the $z$-band attenuation used in SDSS), which originates from the broader distribution in observer-frame $J-K$
color as a function of stellar absorption indices compared to colors at bluer wavelengths such as $r-i$.

\begin{figure}
\epsscale{1.2}
\plotone{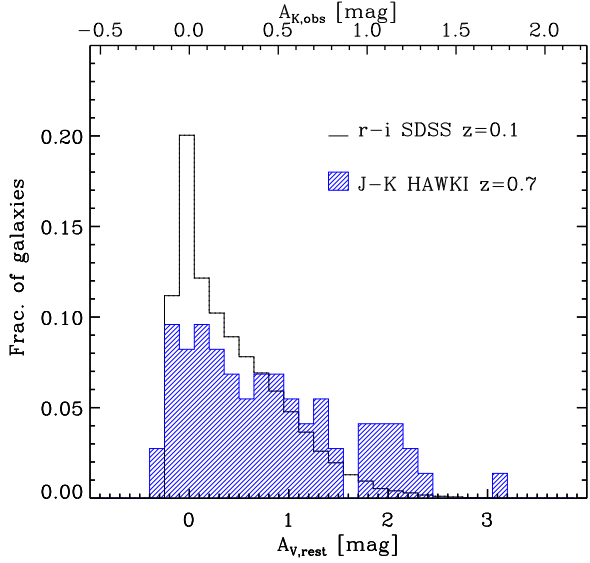}
\caption{Distribution in the dust attenuation inferred from the observer-frame $J-K$ color for our $z=0.7$ sample, converted to rest-frame V-band
attenuation assuming a $\propto\lambda^{-0.7}$ attenuation law (blue hatched histogram). The black histogram shows for comparison the
distribution in $A_V$ estimated from the $r-i$ color excess for a mass-matched SDSS sample. The top x-axis gives the corresponding values of $A_{K,obs}$
which is the quantity directly determined for $z=0.7$ galaxies.}\label{fig:dust_distr}
\end{figure}

Fig.~\ref{fig:dust_distr} shows the distribution in the dust attenuation in the observed K-band, $A_{K,obs}$, inferred from the difference between
the observed $J-K$ color and the color of the dust-free models (see Sec.~\ref{sec:physparam_method}). This is compared to the
distribution of dust attenuation in the observed $z$-band derived for a mass-matched SDSS subsample of galaxies from their
$r-i$ fiber color excess \citep{gallazzi05}. For comparison purposes, both attenuations are converted to $A_V$ assuming a $\propto\lambda^{-0.7}$
attenuation law. The average dust attenuation inferred for our $z=0.7$ sample is $A_V=0.77$ with a scatter of $0.79$. This is larger than the average
$A_V=0.47$ for SDSS galaxies (with a scatter of 0.5~mag). As a further consistency check we have compared the dust attenuation estimates based on
the $J-K$ color excess from our Bayesian fit with those obtained from the continuum fit with \textsc{platefit} (
Sec.~\ref{continuum_fit}) which are based on bluer wavelengths (i.e. the range covered by the IMACS spectra). The two independent estimates compare
well with each other with an average difference of $A_{V,platefit}-A_{V,bayesian}=0.08$ mag with a rms scatter of 0.69 mag, slightly larger than
the typical uncertainty of 0.44 mag in the Bayesian $A_V$ estimates. We also note that, as
expected, galaxies with $A_V>1$ are all detected at 24$\micron$ and classified as star-forming (see Sec.~\ref{sec:sfr_properties}). Conversely, galaxies detected at 24$\micron$ (36/73) have a relatively high
dust attenuation, with an average $A_V$ of 1.36 compared to the average $A_V=0.29$ of non-detections.

\begin{center}
\begin{deluxetable}{lc}
\tablecaption{Range in physical parameters for the model library.}
\tablecolumns{2}
\tablewidth{0pt}
\tablehead{\colhead{Parameter} & \colhead{Range}}
\startdata
Formation time & $\rm 1.5<t_{form}/Gyr< t_{Univ}(z)$  \\  
Exponential timescale & $\rm 0.001 < \gamma/Gyr^{-1} < 1$  \\ 
Mass fraction in a burst & $\rm 0.03< A <4$  \\ 
Duration of a burst & $\rm 3\times10^{7}< t_{burst}/yr<3\times10^{8}$  \\ 
Stellar metallicity & $\rm -1.7<log(Z/Z_\odot)<0.4$  \\ 
Velocity dispersion & $\rm 50<\sigma_V/km/s<350$
\enddata\label{tab:library}
\end{deluxetable}
\end{center}

\begin{center}
\begin{deluxetable*}{lllccccccc}
\tablecaption{Summary of the observed sample.}
\tablecolumns{8}
\tablewidth{18cm}
\tabletypesize{\scriptsize}
\tablehead{\colhead{IMACS ID}& \colhead{COMBO-17 ID} & \colhead{GEMS ID}& \colhead{RA} & \colhead{DEC} & \colhead{$z$} & 
	\colhead{$R_{Vega}$} & \colhead{$f_{24\micron}$}\\
\colhead{(1)} & \colhead{(2)} & \colhead{(3)} & \colhead{(4)} & \colhead{(5)} & \colhead{(6)} & 
\colhead{(7)} & \colhead{(8)}} 

\startdata
obs001     &31712 &GEMSz033220.23m274850.4 & 03:32:20.229 & -27:48:50.42 & 0.73533$\pm$0.00014 & 21.51 &     $<24.6$   \\  
obs002     &23244 &GEMSz033239.48m275301.6 & 03:32:39.488 & -27:53:01.98 & 0.68644$\pm$0.00011 & 20.99 &     $<24.6$   \\ 
obs003     &47365 &GEMSz033228.87m274129.4 & 03:32:28.876 & -27:41:29.17 & 0.73240$\pm$0.00010 & 21.69 &     $<24.6$   \\ 
obs004$^a$ &47568 &GEMSz033237.38m274126.2 & 03:32:37.335 & -27:41:26.10 & 0.66874$\pm$0.00016 & 20.21 &     $<24.6$   \\ 
obs005     &34685 &GEMSz033217.95m274721.3 & 03:32:17.948 & -27:47:21.63 & 0.73290$\pm$0.00012 & 21.30 &     $<24.6$   \\ 
obs006     &29976 &GEMSz033215.97m274943.1 & 03:32:15.973 & -27:49:43.35 & 0.66764$\pm$0.00010 & 20.97 &     $<24.6$   \\ 
obs007     &44584 &GEMSz033209.72m274247.9 & 03:32:09.716 & -27:42:47.79 & 0.73554$\pm$0.00014 & 21.44 &     $<24.6$   \\ 
obs008     &10882 &GEMSz033148.72m275858.1 & 03:31:48.717 & -27:58:58.10 & 0.67844$\pm$0.00010 & 21.36 &     $<24.6$   \\ 
obs010     &18302 &GEMSz033228.05m275526.6 & 03:32:28.049 & -27:55:26.61 & 0.66128$\pm$0.00010 & 21.32 &     $<24.6$   \\ 
obs012     &20786 &GEMSz033246.40m275413.9 & 03:32:46.406 & -27:54:13.91 & 0.66674$\pm$0.00010 & 20.68 &    288.3   \\ 
obs013     &45879 &GEMSz033155.43m274207.5 & 03:31:55.417 & -27:42:07.55 & 0.68143$\pm$0.00010 & 21.16 &    208.0   \\ 
obs014     &27175 &GEMSz033232.73m275102.5 & 03:32:32.737 & -27:51:02.48 & 0.73587$\pm$0.00016 & 21.92 &     $<24.6$   \\ 
obs015     &37979 &GEMSz033144.83m274551.2 & 03:31:44.825 & -27:45:51.38 & 0.73237$\pm$0.00010 & 21.38 &     97.4   \\ 
obs016     &10219 &GEMSz033210.82m275927.0 & 03:32:10.819 & -27:59:27.10 & 0.68188$\pm$0.00010 & 20.61 &     $<24.6$   \\ 
obs017     &23357 &GEMSz033248.18m275256.7 & 03:32:48.178 & -27:52:56.55 & 0.66738$\pm$0.00010 & 20.97 &    163.7   \\ 
obs021     &25642 &		    --     & 03:31:30.569 & -27:51:43.69 & 0.67689$\pm$0.00033 & 21.74 &     $<24.6$   \\ 
obs022     &26418 &		    --     & 03:31:31.743 & -27:51:21.79 & 0.75024$\pm$0.00022 & 21.34 &    207.5   \\ 
obs025     &18411 &GEMSz033300.65m275519.7 & 03:33:00.658 & -27:55:19.86 & 0.70497$\pm$0.00012 & 21.41 &     $<24.6$   \\ 
obs027     &17873 &GEMSz033316.90m275537.1 & 03:33:16.909 & -27:55:37.07 & 0.77318$\pm$0.00018 & 21.39 &     $<24.6$   \\ 
obs029     &34189 &GEMSz033303.80m274736.9 & 03:33:03.805 & -27:47:37.12 & 0.73703$\pm$0.00017 & 21.01 &     57.6   \\ 
obs032     &22031 &GEMSz033321.54m275331.9 & 03:33:21.539 & -27:53:31.97 & 0.77149$\pm$0.00016 & 20.79 &     $<24.6$   \\ 
obs035     &13624 &GEMSz033258.71m275742.7 & 03:32:58.716 & -27:57:42.66 & 0.73578$\pm$0.00012 & 21.36 &     $<24.6$   \\ 
obs036     &20090 &GEMSz033159.89m275431.3 & 03:31:59.901 & -27:54:31.30 & 0.73793$\pm$0.00010 & 21.34 &     58.5   \\ 
obs040     &19631 &GEMSz033205.90m275449.7 & 03:32:05.900 & -27:54:49.65 & 0.68307$\pm$0.00010 & 21.19 &     $<24.6$   \\ 
obs041     &47292 &GEMSz033316.44m274131.6 & 03:33:16.436 & -27:41:31.56 & 0.68212$\pm$0.00010 & 21.03 &     ---    \\ 
obs047     &13869 &GEMSz033142.29m275737.9 & 03:31:42.284 & -27:57:38.08 & 0.73716$\pm$0.00010 & 21.09 &     $<24.6$   \\ 
obs048     &35317 &GEMSz033151.21m274659.3 & 03:31:51.205 & -27:46:59.25 & 0.67358$\pm$0.00010 & 20.76 &    212.6   \\ 
obs049     &32563 &GEMSz033137.40m274830.2 & 03:31:37.395 & -27:48:30.16 & 0.68106$\pm$0.00010 & 20.48 &     $<24.6$   \\ 
obs051     &12969 &GEMSz033135.15m275809.9 & 03:31:35.150 & -27:58:09.84 & 0.67865$\pm$0.00010 & 18.73 &     ---    \\ 
obs052     &13153 &GEMSz033213.73m275753.6 & 03:32:13.726 & -27:57:53.68 & 0.67830$\pm$0.00011 & 21.48 &     $<24.6$   \\ 
obs056     &17809 &GEMSz033202.12m275541.9 & 03:32:02.115 & -27:55:41.97 & 0.67379$\pm$0.00010 & 20.87 &    169.9   \\ 
obs060     &13836 &		    --     & 03:33:13.507 & -27:57:37.66 & 0.68430$\pm$0.00010 & 21.05 &    113.9   \\ 
obs061     &19628 &		    --     & 03:31:55.364 & -27:54:48.11 & 0.73699$\pm$0.00010 & 20.71 &    704.5   \\ 
obs066     &46659 &GEMSz033227.62m274144.9 & 03:32:27.622 & -27:41:44.80 & 0.66528$\pm$0.00015 & 22.25 &    401.7   \\ 
obs074     &19195 &GEMSz033208.42m275455.0 & 03:32:08.419 & -27:54:54.98 & 0.68432$\pm$0.00022 & 22.09 &     $<24.6$   \\ 
obs082     &15578 &GEMSz033144.96m275640.8 & 03:31:44.960 & -27:56:40.83 & 0.71745$\pm$0.00010 & 22.23 &     $<24.6$   \\ 
obs086     &10873 &GEMSz033242.45m275857.6 & 03:32:42.445 & -27:58:57.56 & 0.68577$\pm$0.00010 & 21.89 &     $<24.6$   \\ 
obs096     &46040 &GEMSz033244.65m274202.4 & 03:32:44.644 & -27:42:02.55 & 0.70818$\pm$0.00015 & 22.26 &    397.1   \\ 
obs097     &33358 &GEMSz033235.74m274758.8 & 03:32:35.741 & -27:47:58.86 & 0.66544$\pm$0.00010 & 21.58 &     $<24.6$   \\ 
obs099     &38581 &GEMSz033212.20m274529.9 & 03:32:12.204 & -27:45:29.95 & 0.67769$\pm$0.00010 & 21.79 &     $<24.6$   \\ 
obs114     &42210 &GEMSz033234.34m274350.1 & 03:32:34.346 & -27:43:50.06 & 0.66711$\pm$0.00010 & 21.88 &     $<24.6$   \\ 
obs115     &35480 &GEMSz033221.98m274655.7 & 03:32:21.998 & -27:46:55.78 & 0.66946$\pm$0.00010 & 21.65 &     $<24.6$   \\ 
obs128     &37347 &GEMSz033242.82m274605.6 & 03:32:42.808 & -27:46:05.72 & 0.66891$\pm$0.00016 & 20.99 &    255.3   \\ 
obs135     &44489 &GEMSz033213.86m274248.7 & 03:32:13.855 & -27:42:48.70 & 0.73329$\pm$0.00010 & 21.71 &    333.0   \\ 
obs173$^b$ &32665 &GEMSz033133.31m274818.3 & 03:31:33.314 & -27:48:18.01 & 0.67709$\pm$0.00010 & 21.69 &    219.6   \\ 
obs182     &45775 &GEMSz033157.81m274208.5 & 03:31:57.800 & -27:42:08.51 & 0.66846$\pm$0.00010 & 22.17 &     82.0   \\ 
obs188     &47765 &GEMSz033246.78m274113.9 & 03:32:46.790 & -27:41:14.06 & 0.73188$\pm$0.00011 & 22.08 &     $<24.6$   \\ 
obs199     &39954 &GEMSz033140.61m274449.4 & 03:31:40.616 & -27:44:49.36 & 0.68139$\pm$0.00010 & 22.00 &     $<24.6$   \\ 
obs204     &22401 &GEMSz033325.30m275320.9 & 03:33:25.301 & -27:53:20.86 & 0.70412$\pm$0.00037 & 22.12 &     78.6   \\ 
obs223$^c$ &16885 & 	        --         & 03:31:43.556 & -27:56:03.37 & 0.69990$\pm$0.00010 & 22.22 &     $<24.6$   \\ 
obs232     &11927 &GEMSz033146.94m275826.4 & 03:31:46.944 & -27:58:26.42 & 0.73786$\pm$0.00017 & 22.52 &    176.0   \\ 
obs234     &45243 &GEMSz033320.29m274227.3 & 03:33:20.290 & -27:42:27.30 & 0.68247$\pm$0.00024 & 22.58 &     93.7   \\ 
obs239     &13348 &		    --     & 03:33:03.100 & -27:57:50.69 & 0.68518$\pm$0.00010 & 21.76 &     $<24.6$   \\ 
obs240     &14451 &GEMSz033218.31m275714.4 & 03:32:18.319 & -27:57:14.32 & 0.67220$\pm$0.00022 & 21.98 &     72.2   \\ 
obs245     &46768 &GEMSz033200.55m274143.1 & 03:32:00.549 & -27:41:43.16 & 0.72855$\pm$0.00011 & 22.03 &    161.2   \\ 
obs258     &43803 &GEMSz033239.10m274306.6 & 03:32:39.102 & -27:43:06.56 & 0.66933$\pm$0.00018 & 21.17 &    213.0   \\ 
obs271     &43635 &		    --     & 03:32:59.105 & -27:43:10.70 & 0.68119$\pm$0.00010 & 21.30 &    192.5   \\ 
obs273     &35237 &GEMSz033135.14m274704.5 & 03:31:35.154 & -27:47:04.50 & 0.66326$\pm$0.00019 & 21.28 &    234.4   \\ 
obs294     &14168 &GEMSz033315.87m275724.5 & 03:33:15.860 & -27:57:24.36 & 0.68435$\pm$0.00010 & 21.80 &    167.7   \\ 
obs303     &14166 &		    --     & 03:33:10.110 & -27:57:29.27 & 0.66543$\pm$0.00010 & 21.54 &     $<24.6$   \\ 
obs308     &16550 &		    --     & 03:32:31.074 & -27:56:12.40 & 0.66251$\pm$0.00010 & 22.30 &     $<24.6$   \\ 
obs320     &21347 &GEMSz033238.27m275354.4 & 03:32:38.271 & -27:53:54.39 & 0.68778$\pm$0.00010 & 22.69 &     $<24.6$   \\ 
obs325     &15230 &GEMSz033203.34m275651.8 & 03:32:03.340 & -27:56:51.76 & 0.67833$\pm$0.00010 & 22.15 &    192.1   \\ 
obs326     &44602 &GEMSz033204.93m274242.1 & 03:32:04.926 & -27:42:42.10 & 0.67705$\pm$0.00014 & 22.14 &    113.0   \\ 
obs329     &10466 &GEMSz033236.43m275908.2 & 03:32:36.425 & -27:59:08.24 & 0.66375$\pm$0.00030 & 22.45 &     $<24.6$   \\ 
obs335     &16503 &GEMSz033221.42m275616.9 & 03:32:21.423 & -27:56:17.04 & 0.66043$\pm$0.00010 & 21.87 &    192.6   \\ 
obs338     &41816 &		    --     & 03:32:53.805 & -27:44:00.82 & 0.70918$\pm$0.00026 & 22.11 &     56.5   \\ 
obs349     &16753 &		    --     & 03:31:57.816 & -27:56:04.97 & 0.73355$\pm$0.00021 & 22.36 &    304.0   \\ 
obs381     &20548 &		    --     & 03:31:22.736 & -27:54:12.18 & 0.71089$\pm$0.00022 & 22.22 &     ---    \\ 
obs403     &30682 &GEMSz033127.11m274916.0 & 03:31:27.115 & -27:49:16.04 & 0.68015$\pm$0.00010 & 22.46 &     60.5   \\ 
obs416     &18632 &		    --     & 03:31:39.980 & -27:55:10.92 & 0.67837$\pm$0.00010 & 21.99 &    135.4   \\ 
obs443     &35597 &GEMSz033241.42m274651.5 & 03:32:41.428 & -27:46:51.60 & 0.61984$\pm$0.00019 & 22.40 &     89.0   \\ 
obs444     &38032 &GEMSz033309.36m274546.1 & 03:33:09.354 & -27:45:46.06 & 0.62199$\pm$0.00029 & 21.41 &     $<24.6$   \\ 
obs476     &24917 &		    --     & 03:33:05.457 & -27:52:08.26 & 0.68653$\pm$0.00016 & 22.13 &     54.5   \\ 
obs479     &17808 &		    --     & 03:32:29.438 & -27:55:38.21 & 0.65683$\pm$0.00037 & 20.98 &    521.7   \\ 
obs489     &17052 &GEMSz033209.69m275600.5 & 03:32:09.691 & -27:56:00.22 & 0.73792$\pm$0.00010 & 21.93 &    187.5   \\ 
obs515     &21122 &GEMSz033234.23m275359.9 & 03:32:34.219 & -27:53:59.95 & 0.70432$\pm$0.00021 & 22.26 &     $<24.6$   \\ 
\enddata
\tablecomments{(1): galaxy ID on IMACS mask, (2): galaxy ID in \combo~catalog, (3): galaxy ID in GEMS catalog if available, (4): right-ascension, 
(5): declination, (6): redshift from IMACS spectra, (7): R-band MAGBEST magnitude from 
\combo~catalog, which has formal errors $<0.05$mag, (8): FIDEL Spitzer 24\micron~flux in $\mu$Jy 
when available (non detections are set to 3$\sigma$~limit of 24.6~$\mu$Jy; fluxes of detections are accurate at the 30\% level typically).
(a) The spectrum is contaminated by the light from an object very close and similar in morphology and color to the target. While the inferred stellar population 
properties may be slightly affected, the line broadening and inferred velocity dispersion may be wrong. (b) The spectrum is contaminated by the light of a very blue object 
close in projection but at a different redshift which manifest itself from the presence of emission lines at a redshift of 0.821. (c) The spectrum suffers from a bad 
spectro-photometric calibration because of contamination at red wavelengths.}\label{tab:sample}
\end{deluxetable*}
\end{center}

\section{The stellar populations scaling relations}\label{sec:results}
In the local Universe galaxies are known to follow scaling relations linking the age and the metallicity of their stellar populations
with either their stellar or dynamical mass \citep[see
e.g.][]{gallazzi05,Thomas05,mateus06,graves09a}. In the following sections we explore whether these correlations exist
at intermediate redshift and how they evolve.

\subsection{The stellar age--stellar mass relation}\label{sec:relations_all_age}
Figure~\ref{fig:relations_all}a shows the luminosity-weighted mean age of the $z=0.7$ galaxies in our sample as a function of their stellar mass
(black filled circles). We also show the median and the $16^{th}$ and $84^{th}$ percentiles of the age distribution in bins of stellar mass (blue
filled stars with error bars for galaxies with at least one of the two metal-sensitive indices; panel b). The width of the stellar mass bins 
(0.2~dex) is compatible with the typical uncertainty on stellar mass. When necessary, we increase
the width of the smallest-mass bin in order to include more than three galaxies. Grey empty stars with error bars represent the median and percentiles of
the age distribution in each mass bin for the whole sample, i.e. including also galaxies without \mgfep~and \mgtwofe~(indicated by the grey empty
circles in panel a)

Already at $z=0.7$ there is a trend of increasing light-weighted age with increasing mass, with the most massive galaxies hosting the oldest stellar
populations with an average age $\sim$2.7~Gyr younger than the age of the Universe at that redshift. The stellar age increases on
average from $\sim2.3$~Gyr (1.5~Gyr including all galaxies) around $3\times10^{10}M_\odot$ to $\sim4.5$~Gyr over about an order of
magnitude in stellar mass. This relation can be directly compared to the one obtained for a much larger sample of SDSS galaxies at a
median redshift $z=0.1$ in \cite{gallazzi05}, reported in Fig.~\ref{fig:relations_all}b with red solid line (median) and shaded region (percentiles). 
It is immediately clear that $z=0.7$
galaxies are significantly younger than their local counterparts. A difference of about 0.4~dex (4.5~Gyr, close to the evolution in
the age of the Universe) is observed between the upper envelopes of the two distributions at a mean stellar mass of $10^{11}M_\odot$.
Interestingly the absolute increase in age with mass observed at $z=0.7$ is more modest compared to the average increase of 3.4~Gyr for
low-redshift galaxies between $3\times10^{10}$ and $3\times10^{11}M_\odot$. This indicates that the rate of evolution in light-weighted age with redshift 
is a function of stellar mass with less massive galaxies experiencing a smaller age change, probably associated to a more prolonged star formation activity between $z=0.7$ 
and today and/or a larger fraction of star-forming galaxies at lower masses. We come back to this point below and in Sec.~\ref{sec:discussion2}.

Panel (c) of Fig.~\ref{fig:relations_all} compares the $\rm 84^{th}-16^{th}$ percentile range in the distribution of log light-weighted age at fixed stellar mass (diamonds) with the
typical uncertainty in age (crosses) for our $z=0.7$ sample (filled symbols) and for SDSS galaxies (empty symbols).  In both samples the
inter-percentile range of the age distribution increases to lower stellar masses and it is always significantly larger than the
uncertainties. At low redshifts the largest dispersion occurs at masses below $3\times10^{10}M_\odot$, the characteristic transition
regime where galaxies of different morphology and star formation activity mostly overlap \citep[e.g.][]{kauffmann03,gallazzi05}. At
$z=0.7$ our data only probe stellar masses higher than this threshold where we observe a similar scatter in $\log(t_r/yr)$ as
locally. 

To further quantify the redshift evolution of the ages of massive galaxies we fit the median values as a
function of mass adopting a slightly modified version of the functional form proposed by \cite{moustakas11} to describe the relation between stellar
mass and gas-phase metallicity:
\begin{equation}
P(M_\star) = \bar{P} - \log \frac{\left(1+\left(\frac{\bar{M}}{M_\star}\right)^\gamma\right)}{\left(1+\left(\frac{\bar{M}}{10^{11.5}M_\odot}\right)^\gamma\right)}
\end{equation} \label{eqn:fit}
where P in this case is the luminosity-weighted age, $\log(t_r/yr)$. With the normalization adopted here, the parameter $\bar{P}$
represents the characteristic age at a stellar mass of $10^{11.5}M_\odot$. The parameter $\bar{M}$ is the characteristic mass at which
the relation bends, while $\gamma$ governs the slope of the transition: for $\gamma\gg1$ the relation is flat above $\bar{M}$, while
for $\gamma\ll1$ it remains steep even above $\bar{M}$.
The fitted curve to the $z=0.7$ median ages is shown with a blue solid line in Fig.~\ref{fig:relations_all}a,b, where it can be seen
that the proposed form describes well the median trend in age. The grey dashed curve shows the fit to the full sample: including also
galaxies for which neither of the two metallicity-sensitive indices can be measured produces only a mild steepening of the age--mass
relation at masses $\log(M_\star/M_\odot)\lesssim10.8$. We note that, as discussed in Appendix~\ref{sec:physparam_systematics_indx}, there is
no indication of a bias in the age estimates of these galaxies, but we show the age--mass relation obtained both including and
excluding them for consistency with the metallicity--mass relation. For comparison purposes we fit the same analytic form to the
SDSS median ages provided in \cite{gallazzi05} (the analytic fit overlaps almost perfectly with the median trend and is thus not
displayed in the figure). The median relations at $z=0.7$ are given in Table~\ref{tab:median_relation} and the parameters of the best fit curve for 
both $z=0.7$ and $z=0.1$ (SDSS) are summarized in Table~\ref{tab:curve_fit}. 

We find that at $z=0.7$ the characteristic age at $10^{11.5}M_\odot$ is $\log(t_r/yr)=9.66\pm0.04$, i.e. $\sim$4~Gyr younger than the present-day
value. This difference is 1~Gyr smaller than the time elapsed between $z=0.7$ and $z=0.1$. The parameter $\gamma$ ($1.09\pm0.98$) is consistent
within the uncertainties with the value fitted on the local relation. The characteristic mass $\bar{M}$ is instead about twice as large as the local
value. From a $\chi^2$ test we find that this difference between $z=0.7$ and $z=0.1$ is significant only when including all the galaxies in the sample
(dashed curve). Moreover, we caution that selection effects may limit our sensitivity to lower values of the transition mass as we lose completeness of
the quiescent population around masses of $\rm10^{10.5}M_\odot$.
It is however interesting to note that at both redshifts the estimated characteristic mass at which the relation starts to flatten is similar to
the so-called transition mass at which the mass functions of the star-forming and quiescent galaxy populations cross \citep[e.g.][]{bell07}. The
difference we find in characteristic mass between $z=0.7$ and $z=0.1$ resembles the slight increase with redshift of the transition mass
\citep[e.g.][]{pannella09,muzzin13}.

The cyan dotted curve in Fig.~\ref{fig:relations_all}b displays the relation obtained by subtracting from the local relation the look-back time between
$z=0.7$ and $z=0.1$. This corresponds to the age evolution of an SSP, hence the simplest form of passive evolution.
There are two main differences with respect to the observed
relation at $z=0.7$: i) the observed characteristic light-weighted age (i.e. the age at $10^{11.5}M_\odot$) is 0.1~dex (i.e $2\sigma$) older than
predicted and ii) the decrease in age with decreasing mass is significantly shallower than predicted. In
other words, this comparison indicates that simple passive evolution of the observed $z=0.7$ galaxy population does not correctly predict the local
relation, in that the $z=0.7$ galaxies would only make up the oldest portion of the present-day galaxy distribution. Moreover, the rate of evolution
is mass dependent. We note, however, that the passive evolution of the light-weighted (or mass-weighted) mean age of a complex stellar population is not simply given by
the elapsed time, and we address this point later.

\subsection{The stellar metallicity--stellar mass relation}\label{sec:relations_all_zstar}
\begin{figure*}
\epsscale{1.1}
\plotone{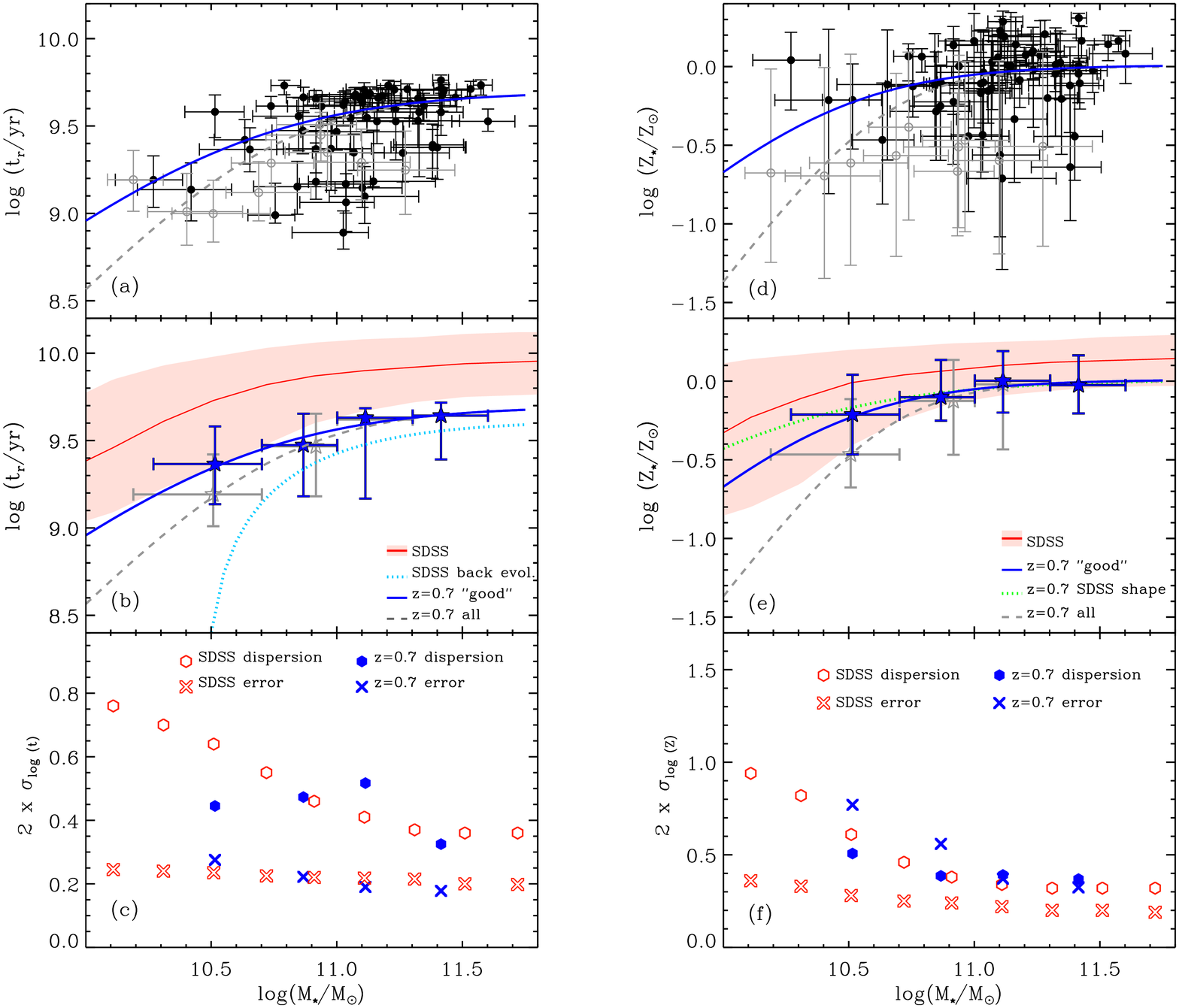}
\caption{{\it Upper panels:} Mean luminosity-weighted age ({\it a}) and stellar metallicity ({\it d}) versus stellar mass for the $z=0.7$ galaxies in our sample
(the grey empty circles indicate those galaxies with no metallicity-sensitive indices available). For each galaxy we plot the median-likelihood estimate of the
parameters, with error bar given by the $16^{th}-84^{th}$ percentile range of the PDF. {\it Middle panels:} Filled stars show the median stellar age ({\it b}) and
stellar metallicity ({\it e}) in bins of stellar mass 0.2~dex wide (or larger in order to include more  than 3 galaxies), and the associated error bars show the
$16^{th}$ and $84^{th}$ percentiles of the distribution (the median parameters in bins of mass are given in Table~\ref{tab:median_relation}). The solid curve 
shows the fit to the median parameter values adopting the functional form of
Eq.~\ref{eqn:fit} (the fitted parameters are summarized in Table~\ref{tab:curve_fit}). The grey empty stars and dashed line show the corresponding results when
including the whole sample. These curves are repeated in panels {\it a} and {\it d}, respectively. In both panels {\it b} and {\it e} the red solid line and colored
region show the corresponding median relations and dispersion for SDSS galaxies as  derived in \cite{gallazzi05}. The dotted curve in panel {\it b}  shows the
age--mass relation expected at $z=0.7$ by passively evolving the $z=0$ SDSS relation. In panel {\it e} the dotted curve shows the fit obtained fixing the shape of
the metallicity--mass relation to the local one.   {\it Lower panels:} The filled diamonds show the $\rm84^{th}-16^{th}$ percentile range in stellar age ({\it c})
and stellar metallicity ({\it f}) in bins of stellar mass in comparison to the median uncertainty in the corresponding parameter, for the z=0.7 `good' sample (filled
crosses). The empty symbols refer to the same quantities for SDSS high-S/N galaxies as in \cite{gallazzi05}.}\label{fig:relations_all}
\end{figure*}

Figure~\ref{fig:relations_all}d shows the distribution of stellar metallicity as a function of stellar mass for our sample at $z=0.7$
(black filled circles; empty grey circles indicate galaxies without a measure of the Mg-Fe composite indices, for which the
metallicity estimate is thus highly uncertain). The median and percentiles (blue filled stars with error bars) of the metallicity
distribution in the same mass bins as in Fig.~\ref{fig:relations_all}b are shown in Fig.~\ref{fig:relations_all}e. Above $3\times10^{10}M_\odot$, the $z=0.7$ galaxies display a
significant scatter in metallicity but the median values show a mild increase with stellar mass of about 0.2~dex over an order of
magnitude in mass. For comparison, the red solid line and shaded region show the median and percentiles of the local relation obtained for
SDSS galaxies \citep{gallazzi05}. Also for low-redshift galaxies the correlation between stellar metallicity and stellar mass becomes
significantly flatter above $3\times10^{10}M_\odot$. 
Although many $z=0.7$ galaxies occupy the same locus as the local population, it is notable that the median metallicity at $z=0.7$ is
systematically lower than the local one. In particular, the distribution is broadened toward lower metallicities
with respect to the $z=0$ distribution. However the scatter at $z=0.7$, as measured by the $16^{th}$ and $84^{th}$ percentiles, is
comparable to the uncertainties in stellar metallicity at $z=0.7$, as can be seen in Fig.~\ref{fig:relations_all}f. If we considered the full sample the median metallicity at masses above
$10^{11}M_\odot$ would be unchanged, but we would find a stronger decrease in metallicity with stellar mass below $10^{11}M_\odot$
(empty grey stars). As discussed in Appendix~\ref{sec:physparam_systematics_indx}, excluding the Mg-Fe composite indices from the fit results
in systematically lower metallicities by 0.2~dex. However, the galaxies that lack a measure of the Mg-Fe indices are young galaxies
which probably have intrinsically weak metal absorptions (thus hard to measure in low-S/N spectra) and hence low metallicity. 
If this is the case the true median metallicity at $10^{10.5}M_\odot$ might be lower than that shown by the blue filled stars.

To be more quantitative, we fit the function in Eqn.~\ref{eqn:fit} to the median metallicities and masses at $z=0.7$ and the result
is shown by the solid blue line for 'good' data only and by the grey dashed line for the sample as a whole. The same analytic form is fit to the
median trend obtained for SDSS galaxies at $z\sim0.1$ (the fit is not shown but would lie on top of the median relation in
Fig.~\ref{fig:relations_all}e)\footnote{We stress that the functional description as in Eqn.~\ref{eqn:fit} of the relations of age
and metallicity with stellar mass is valid only over the mass range probed in this work. For SDSS, in particular, we know that it
deviates significantly from the median trends in \cite{gallazzi05} below $10^{10}M_\odot$.}. The $z=0.7$ characteristic metallicity
at $10^{11.5}M_\odot$ is $\log(Z_\star/Z_\odot)=0.0\pm0.04$ (regardless of whether the whole sample is used or not), i.e.
$0.13\pm0.04$~dex lower than the local value. The shape of the relation, as expressed by the parameters $\bar{M}$ and $\gamma$, is
instead consistent with the local one. This is clearly shown by the green dotted line which is the fit obtained by fixing $\bar{M}$ and
$\gamma$ to the SDSS values. As noted above, by including all galaxies we would fit a steeper relation at masses below
$10^{11}M_\odot$. This suggests a stronger chemical enrichment for less massive galaxies since $z=0.7$.

\begin{center}
\begin{deluxetable}{ccc}
\tablecaption{Median ages and metallicities at $z=0.7$ in bins of stellar mass}
\tablecolumns{3}
\tablewidth{8cm} 
\tablehead{\colhead{$\rm \log(M_\ast/M_\odot)$} & \colhead{$\rm \log(Z_\ast/Z_\odot)$} & \colhead{$\rm \log(t_r/yr)$}}
\startdata
\cutinhead{"good" data}
$11.41 (11.30,11.60)$ & $-0.02 (-0.20,0.16)$ & $9.64 (9.39,9.72)$ \\
$11.11 (11.00,11.30)$ & $ 0.00 (-0.20,0.19)$ & $9.63 (9.17,9.68)$ \\ 
$10.87 (10.70,11.00)$ & $-0.10 (-0.25,0.13)$ & $9.47 (9.18,9.65)$ \\
$10.51 (10.27,10.70)$ & $-0.21 (-0.47,0.04)$ & $9.37 (9.14,9.58)$ \\
\cutinhead{all data}
$11.41 (11.30,11.60)$ & $-0.02 (-0.20, 0.16)$ & $9.64 (9.39,9.72)$ \\
$11.11 (11.00,11.30)$ & $-0.02 (-0.43, 0.19)$ & $9.62 (9.17,9.68)$ \\ 
$10.92 (10.70,11.00)$ & $-0.13 (-0.47, 0.13)$ & $9.47 (9.18,9.65)$ \\
$10.51 (10.19,10.70)$ & $-0.47 (-0.68,-0.11)$ & $9.19 (9.01,9.42)$ \\
\enddata\label{tab:median_relation}
\tablecomments{First column: median stellar mass of each bin; the numbers in parenthesis give the 
minimum and maximum mass of the bin. Second (third) column: median stellar metallicity (age) in bins of 
stellar mass; the numbers in parenthesis give the $16^{th}$ and $84^{th}$ percentiles of the distribution. 
The upper set of rows refers to our more robust estimates based on only good quality data (filled stars in Fig.~\ref{fig:relations_all}b,e); the lower set 
of rows refers to the results on the whole sample (empty stars in Fig.~\ref{fig:relations_all}b,e).}
\end{deluxetable}
\end{center}

\begin{center}
\begin{deluxetable}{lccc}
\tablecaption{Polynomial fit to the stellar populations scaling relations}
\tablecolumns{4}
\tablewidth{8cm}
\tablehead{\colhead{redshift} & \colhead{$\bar{P}$} & \colhead{$\bar{M}/10^{10}M_\odot$} & \colhead{$\gamma$}}
\startdata
\cutinhead{$\log(t_r/yr)$}
$z=0.7$, "good" & $9.66\pm0.04$  & $4.0\pm2.5$ & $1.09\pm0.98$ \\
$z=0.7$, "all"  & $9.66\pm0.04$  & $5.6\pm1.9$ & $1.45\pm0.57$ \\
$z=0.1$ & $9.94\pm0.001$ & $2.44\pm0.04$ & $1.19\pm0.02$ \\

\cutinhead{$\log(Z_\star/Z_\odot)$}
$z=0.7$, "good" & $0.00\pm0.04$ & $2.6\pm1.0$ & $1.4\pm1.2$ \\
$z=0.7$, "good"\tablenotemark{b} & $-0.01\pm0.02$ & $1.7$ & $1.05$ \\
$z=0.7$, "all" & $0.00\pm0.04$ & $4.5\pm0.8$ & $2.1\pm0.7$ \\
$z=0.1$ & $0.13\pm0.006$ & $1.7\pm0.16$ & $1.05\pm0.15$ \\
\enddata\label{tab:curve_fit}
\tablecomments{Fit to the median ages and metallicities as a function of stellar mass for our IMACS $z=0.7$ sample, either considering the whole sample ("all", empty
symbols in Fig.~\ref{fig:relations_all}b,e) or only galaxies for which at least \mgtwofe~or \mgfep~
is available ("good", filled symbols in Fig.~\ref{fig:relations_all}b,e), and for the SDSS sample ($z=0.1$). We adopt the functional form
$P=\bar{P} +\log(1+(\bar{M}/10^{11.5}M_\odot)^\gamma) -
\log(1+(\bar{M}/M_\star)^\gamma)$ where the parameter P is either the luminosity-weighted age $\log(t_r/yr)$ or the
stellar metallicity $\log(Z_\star/Z_\odot)$. We stress that the fit to the SDSS relations is valid only in the mass range
considered here, i.e. above $10^{10}M_\odot$, but it is not a good representation of the local relations below this range.}
\tablenotetext{b}{Only the zero-point $\bar{P}$ of the relation is being fitted, while the
parameters $\bar{M}$ and $\gamma$ are kept fixed to the $z=0.1$ values. This fit is shown as green dotted curve in
Fig.~\ref{fig:relations_all}e.}
\end{deluxetable}
\end{center}

\section{The dependence on star formation activity}\label{sec:relations_ssfr}
In this section we explore the star formation activity of the galaxies in our sample and how it affects the location of
galaxies in the stellar population scaling relations discussed in the previous sections and their expected evolution.
\subsection{Star formation properties}\label{sec:sfr_properties}
Estimates of SFR are obtained by combining the UV luminosity, estimated from the \combo~synthetic band centered at 2800\AA,
and the IR luminosity, estimated from the Spitzer 24\micron~flux, adopting the calibration described in \cite{gallazzi09}
converted to a Chabrier IMF. There are 36/73 galaxies in our sample with a 3$\sigma$ detection at 24\micron, while 34 galaxies have only
a 3$\sigma$ upper limit flux of 24.6 $\mu$Jy (the remaining 3 galaxies are not covered by MIPS observations).
For galaxies not detected at 24\micron~or not covered by MIPS observations we include
only the UV term in the SFR estimate, which then should be regarded as a lower limit to the true SFR. The typical uncertainty on the 24\micron~flux for the detections
in our sample is $\sim$30\%. Conservative error estimates on the SFR amount to $\sim$0.3 dex \citep{bell05}.

\begin{figure}
\epsscale{1.1}
\plotone{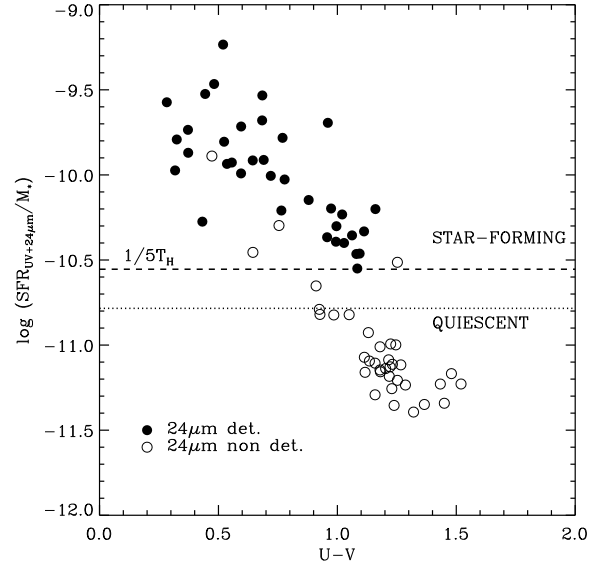}
\caption{Specific SFR versus rest-frame $U-V$ color. Filled circles are galaxies with a $3\sigma$ detection at
24\micron, empty circles are galaxies with only upper limits at 24\micron~(or no MIPS observations) for which the SFR is estimated from UV
only. The dashed line marks the specific SFR that corresponds to a timescale of 5 times the age of the
Universe at $z=0.7$: we consider galaxies below this line as quiescent. The dotted line indicates the value used later at $z=0.1$ to classify 
quiescent and star-forming galaxies.}\label{fig:ssfr_col}
\end{figure}

Despite the uncertainties associated with the derivation of total UV and IR luminosities from monochromatic fluxes and with their
calibration into SFR estimates, we verified the consistency of these SFR estimates with other indicators of galaxy star formation
activity. 
We first looked at the distribution of our sample in specific SFR (the SFR divided by the stellar mass obtained in
Sec.~\ref{sec:physparam_method}) against the rest-frame $U-V$ color from \combo~ in
Fig.~\ref{fig:ssfr_col}, which shows a nice correlation between the two quantities. Galaxies not detected at 24\micron~(empty circles) 
have typically red colors (thus unlikely to be associated with dust attenuation) and low specific SFR. The dashed line
indicates the value of specific SFR that corresponds to a timescale of 5 times the age of the Universe at $z=0.7$.
Virtually all the galaxies detected at 24\micron~are forming stars at a faster rate, while non detections
are mostly in a quiescent phase. The few blue galaxies not detected at 24\micron~have a specific SFR higher than 1/(5T$_H$)
and would thus be classified as star forming even if their SFR were underestimated. In what follows we define as quiescent
(star-forming) those galaxies with a specific SFR lower (higher) than 1/(5T$_H$).

\begin{figure}
\epsscale{1.1}
\plotone{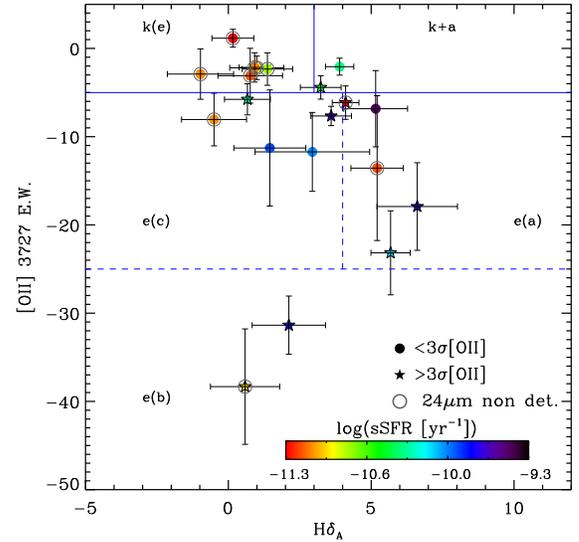}
\caption{Diagnostic diagram of recent star formation mode showing the equivalent width of \oii~3727 in emission (we adopt the convention of negative values of EW for
emission lines) and the strength of the \hda~absorption. We exclude from this plot galaxies which do not have spectral coverage in \oii. Stars (circles) indicate
galaxies with a detection of \oii~emission above (below) 3$\sigma$. The color of the symbols reflect the specific SFR (from UV+24\micron). The grey empty circles
indicate galaxies with only upper limits at 24\micron~and for which the SFR contains only the UV term. Solid and dashed lines indicate the ranges in \oii~3727
and \hda~of the  \cite{poggianti09} classification: passively evolving galaxies (k), post-starburst galaxies (k+a), quiescent star forming (e(c)), starbursts (e(b))
and dusty starburst candidates (e(a)).}\label{fig:oii_hdelta}
\end{figure}

Fig.~\ref{fig:oii_hdelta} shows the distribution in \oii~3727 emission and \hda~absorption index for the 20 galaxies in our
sample with spectral coverage in \oii. Galaxies with a $>3\sigma$ detection of \oii~emission are shown as stars. The color of the symbols
is associated to the galaxy specific SFR 
(encircled points identify galaxies with only an upper limit on the 24\micron~flux). This diagram is a diagnostic of the current and
recent-past star formation activity \citep{Dressler83}. The classification of
\cite{poggianti09} is indicated in Fig.~\ref{fig:oii_hdelta}: passively evolving galaxies (k), post-starburst galaxies (k+a), quiescent star forming (e(c)), starbursts (e(b))
and dusty starburst candidates (e(a)). 
The figure shows that for galaxies detected at 24\micron~there is a satisfactory correspondence between specific SFR and \oii~EW. Galaxies not
detected at 24\micron~have specific SFR below $1.6\times10^{-11} yr^{-1}$ and the majority of them have low or undetected \oii~emission consistent
with the location of currently passive galaxies. Among this subsample with \oii~spectral coverage there are only two objects (i.e. 10\%) with an EW
stronger than $-10$\AA~but with low specific SFR from UV alone. The one with the strongest \oii~3727 ($\rm EW=-38.3$, obs007) has a low dust attenuation of
$A_V=0.04$, according to our analysis of absorption indices and $J-K$ color, and thus may have indeed negligibly dust-obscured star formation. The
other one ($\rm EW=-13.6$, obs003) has a higher $A_V$ of 0.9~mag, thus its SFR(UV) may be underestimated by a factor of 3-4, which is still not sufficient to bring
this galaxy above our quiescent threshold. It should be noted, however, that \oii~emission
has beed detected in more than 50\% of SDSS red-sequence galaxies. Most of the \oii-emitting red-sequence galaxies show a high \oii/\ha ratio and
emission line ratios typical of LINERs \citep[e.g.][]{Yan06}, hence their \oii~emission is not associated to star formation. We cannot verify the
emission line ratios for our sample since our spectra do not extend beyond $\sim5500$\AA, but this could in principle explain the high
\oii~EW of those galaxies in Fig~\ref{fig:oii_hdelta} with low specific SFR (note, though, that their \oii~EW are higher than those
typical of \cite{Yan06} red-sequence galaxies).

It is also interesting to note that, according to this diagnostic diagram, we identify three galaxies (obs001, obs029, obs036) consistent within
their error bars with the post-starburst location. A fourth one (obs489) would be consistent with the k+a position due to the relatively large
error on \oii~3727, but its specific SFR (from UV and 24\micron) is among the highest in our sample ($\log(sSFR)=-9.5$). We estimate a relatively small dust
attenuation of $A_V=0.56^{+0.36}_{-0.46}$ mag, thus likely not significantly affecting the \oii~emission. Inspecting the GEMS HST images of this 
galaxy we find that there is a nearby faint object likely contaminating the Spitzer 24\micron~flux, thus potentially leading to a slight overestimate of the SFR. We note that, 
even assuming the 24\micron~upper limit, obs489 would still be classified as star-forming with a specific SFR of $\log(sSFR)=-10.23$. It could also be that the total
\oii~3727 emission is underestimated due to flux missed by our 1"-wide slit. Indeed, the bluest (most star forming) regions are in the outer part of this galaxy.

Assuming that galaxies with \oii~coverage are a random subset of the whole sample we would expect an overall
fraction of $15\pm8$\% of post-starburst galaxies. We note though that the classification of post-starburst galaxies based on their \oii~emission
is uncertain for two reasons. The \oii3727 line may be undetected because attenuated by dust, thus leading to a potential overestimate of the
fraction of post-starburst galaxies. Indeed two of the post-starburst candidates identified in Fig.~\ref{fig:oii_hdelta} are detected at
24\micron~and have a specific SFR slightly higher than our threshold for quiescent galaxies. On the other hand, if \oii~emission does not entirely
come from star formation, as discussed above, also the number of post-starburst galaxies (identified by the lack of \oii) could be underestimated. 
To bypass both of these problems, we can directly use our estimates of total SFR and select post-starburst candidates as those galaxies with
specific SFR lower than $1/(5t_H)$ and with \hda$>3$. We find that only 2 galaxies in the whole sample (obs001, obs003) satisfy this criterion,
giving a post-starburst fraction of only $3\pm2$\%.

\subsection{The dependence of age and metallicity scaling relations on specific star formation rate}\label{relations_ssfr}
\begin{figure} 
\epsscale{1.2}
\plotone{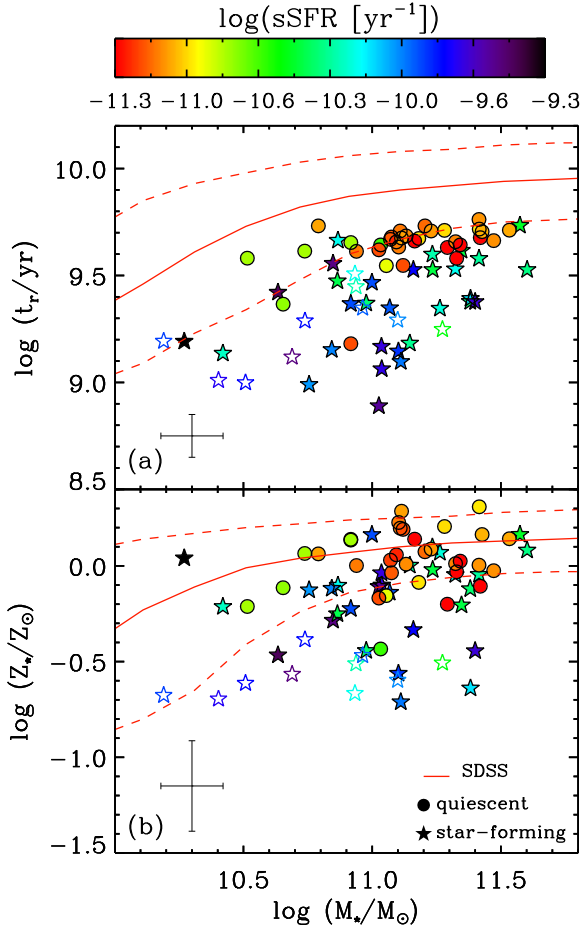}
\caption{Luminosity-weighted mean age (a) and stellar metallicity (b) as a function of stellar mass for galaxies at $z=0.7$. The color of the
symbols is associated to the specific SFR of each galaxy, where the SFR is based on a combination of UV and 24\micron~ for galaxies
detected in Spitzer and from UV only for galaxies with no detection at 24\micron~ or not covered by Spitzer/MIPS.
Galaxies classified as star-forming according to Fig.~\ref{fig:ssfr_col} are shown as stars, while quiescent galaxies are shown as circles.
Hollow symbols indicate galaxies with poorer quality spectra for which a measure of the metal-sensitive
indices could not be obtained and hence whose metallicity estimate is more uncertain. The error bar in the bottom left corner shows the average
uncertainty on the physical parameters. As a reference the corresponding relations for low-redshift galaxies from SDSS are shown with red
solid (median) and dashed curves ($16^{th}$ and $84^{th}$ percentiles).}\label{fig:relations_ssfr}
\end{figure}

In Fig.~\ref{fig:relations_ssfr} we explore whether and how the location of galaxies in the age--mass (panel a) and metallicity--mass planes
(panel b) depends on their star formation activity and how this can affect their expected evolution. The $z=0.7$ galaxies are represented
with colored symbols that reflect their specific SFR. Galaxies classified as quiescent according to Fig.~\ref{fig:ssfr_col}
are shown with circles, while star-forming galaxies are shown with stars. Hollow symbols refer to objects with poorer quality spectra for which a measure of 
the composite Mg-Fe indices was not achieved. We repeat the corresponding SDSS relations for reference (red lines). It is immediately
clear that differences in the galaxy star formation activity play a significant role in the scatter of both age and stellar metallicity at
fixed mass. Quiescent galaxies with low specific SFR are predominantly found at masses above $10^{11}M_\odot$, they are the oldest galaxies
at $z=0.7$ with a light-weighted age of $\log(t_r/yr)\sim9.7$ (i.e. about 2~Gyr younger than the Universe age) and a stellar metallicity typically above
solar, consistent with local massive galaxies. Star-forming galaxies span a large range in stellar mass (the whole range probed
by our sample) but are clearly offset in the stellar age--mass relation towards younger ages with respect to quiescent galaxies, as naturally expected from their ongoing star
formation. They also
extend to lower metallicities than quiescent galaxies contributing to the scatter around a metallicity value of 30\% solar at all masses and in particular around
$10^{11}M_\odot$. However we note that some of the more actively star forming galaxies have already stellar metallicities comparable to quiescent galaxies at
$z=0.7$ and to local SDSS galaxies.

The star formation rate has been shown to play an important role as a second variable in the gas-phase metallicity versus mass relation: throughout the cosmic
time (up to $z\sim2.5$) the scatter around the mass-metallicity relation at given M$_\star$ correlates with SFR, with lower SFR being associated to higher
gas-phase metallicities \citep{Mannucci10,LaraLopez10}. Our $z=0.7$ sample, including both star-forming and quiescent galaxies, shows a qualitatively similar
trend such that at fixed mass galaxies with low (specific) SFR tend to have higher metallicities than more actively star-forming galaxies.
However if we consider only star-forming galaxies we cannot statistically confirm
the existence of a relation of stellar metallicity with both specific SFR and mass, due to the relatively small number of galaxies and the uncertainties on stellar metallicity for
star-forming galaxies. We also find a negative correlation between the residuals in light-weighted age from the median relation and the specific SFR such that
galaxies with higher specific SFR deviate to younger ages at fixed mass. Such a trend is evident also if we consider star-forming galaxies only.

\begin{figure}
\epsscale{1.3}
\plotone{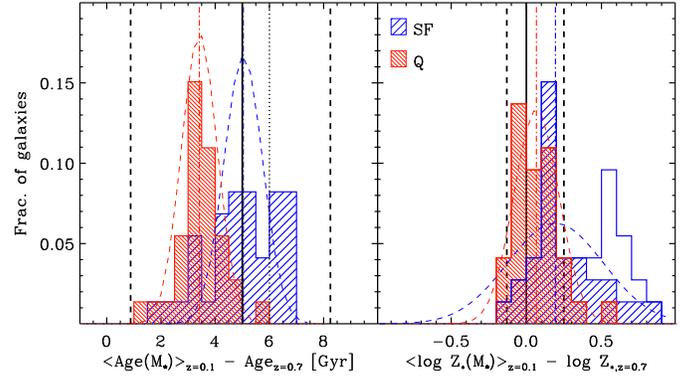}
\caption{{\it Left panel:} Distance of the light-weighted ages of $z=0.7$ galaxies from the median local relation obtained from SDSS for all galaxy types. The solid vertical line 
shows the look-back time between $z=0.7$ and $z=0.1$ (5 Gyr), which corresponds to the passive evolution of an SSP. The dashed lines show the minimum
and maximum offset allowed for passively evolved $z=0.7$ galaxies to be
compatible with, respectively, the $84^{th}$ and $16^{th}$ percentiles of the local age distribution. The dotted vertical line indicates the age evolution 
expected from passive evolution of composite stellar populations (6 Gyr). The red hatched histogram refers to quiescent galaxies, while the blue hatched histogram
to star-forming galaxies. 
{\it Right panel:}  Distance of the stellar metallicities of $z=0.7$ galaxies from the median local relation obtained from SDSS for all galaxy types. The solid
vertical line shows no
evolution in metallicity, while the dashed vertical lines show the minimum and maximum offsets allowed for the metallicity of $z=0.7$ galaxies to be
consistent with the $84^{th}$ and $16^{th}$ percentiles of the present-day metallicity distribution. The red hatched histogram refers to quiescent galaxies, the blue empty histogram refers to all
star-forming galaxies, while the blue hatched histogram to only star-forming galaxies with a measure of metal-sensitive indices. 
The dashed curves show gaussians
centered at the median offset (dot-dashed lines) with a width given by the median error on the parameter.}\label{fig:distr_offset}
\end{figure}

To better visualize the comparison between massive galaxies at $z=0.7$ and the local ones, we plot in Fig.~\ref{fig:distr_offset} the
difference in age (left panel) and metallicity (right panel) between the local relations and the individual $z=0.7$ galaxies, for quiescent galaxies (red
histograms) and star-forming galaxies (blue histograms). As a reference, the solid vertical
line in the left panel indicates the time elapsed between $z=0.7$ and $z=0.1$ (the average redshift of the
SDSS sample). This represents the simplest form of passive evolution, i.e. the age evolution of a simple stellar population. 
Galaxies with a distance from the local relation equal to this value would end up on the local
relation if they evolved passively since $z=0.7$. Passive evolution would bring $z=0.7$ galaxies into agreement with the
upper (lower) inter-percentile range of the local distribution if their distance to the local relation were consistent with the
lower (upper) dashed vertical lines. The dotted vertical line indicates an age difference of 6 Gyr predicted for the passive evolution of composite 
stellar populations (see Sec.~\ref{sec:discussion2}), which is {\it larger} than that of an SSP. 

Virtually all quiescent galaxies are offset from the local relation by less than the look-back time, being on average 3.4 Gyr younger than the median age of
present-day equally massive galaxies. Star-forming galaxies have a median offset from the local relation consistent within the average error on age with the time elapsed between
$z=0.7$ and $z=0.1$. If $z=0.7$ star-forming galaxies with $M_\star>10^{11}M_\odot$ had their star formation rapidly quenched and then they evolved
passively until the present, they would contribute to the $\rm16^{th}-50^{th}$ inter-percentile range of the local massive galaxy population. At masses above
$10^{11}M_\sun$ our $z=0.7$ sample is made up by $62\pm11$\% of quiescent galaxies and $38\pm9$\% of star-forming galaxies in number. However, as also shown in
Fig.~\ref{fig:relations_all}b, even if this population of $z=0.7$ galaxies would evolve to ages consistent with local galaxies, under passive evolution hypothesis, it is not
sufficient to reproduce the median trend at $z=0$. This implies that 1) passive
evolution is not viable already at masses $>10^{11}M_\odot$ and that at least some of the $z=0.7$ massive galaxies experience new or
continued star formation and hence their light-weighted ages evolve less than the time between $z=0.7$ and $z=0.1$ or/and 2) lower-mass
star-forming galaxies need to evolve non-passively and grow in mass to further populate the young portion of massive
local galaxies.

The right-hand panel of Fig.~\ref{fig:distr_offset} shows that the stellar metallicities of quiescent galaxies at $z=0.7$ are consistent with the distribution of
present-day galaxies: the offsets in stellar metallicity of quiescent galaxies peak around zero, but the distribution is slightly skewed to positive offsets with a
median at 0.07~dex. Star-forming galaxies have a median offset of 0.19~dex, with a tail to larger offsets (not entirely associated to galaxies with poorer quality
data). At face value, a fraction of $44\pm16$\% ($53\pm12$\% if we include the poorer quality spectra) of the $z=0.7$ star-forming galaxies
have a metallicity lower than the $\rm16^{th}$ percentile of the local distribution (but consistent with it within the errors). Interestingly, these galaxies have high specific SFR (see
Fig.~\ref{fig:relations_ssfr}b).

\subsection{Quiescent versus star-forming galaxies}\label{sec:relations_passive}
\begin{figure*}
\epsscale{1}
\plotone{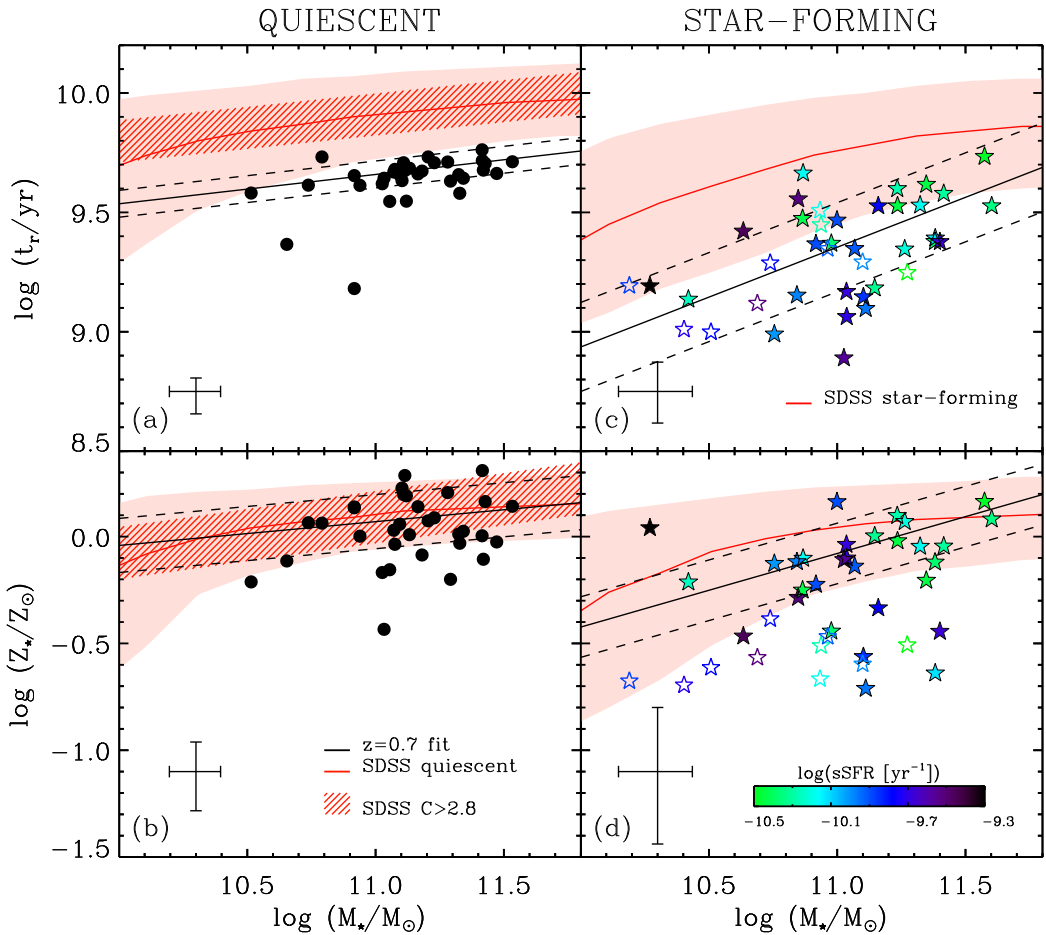}
\caption{Luminosity-weighted mean age (panels a,c) and stellar metallicity (panels b,d) as a function of stellar mass for galaxies at $z=0.7$ separated into passive
(left panels) and actively star forming (right panels) based on a cut in specific SFR at $1/(5 t_H)$. Star-forming galaxies are color-coded on the basis of their specific SFR. 
The black lines in each panel indicate a linear regression fit
to the `good' data only (filled symbols), weighted by the errors on individual measurements and the associated scatter. The red curve and shaded area in each panel display the corresponding  median relation
for SDSS galaxies separated into quiescent and star-forming using the same criterion on specific SFR as at $z=0.7$.  The red hatched region in the left-hand panels
shows the linear fit of \cite{gallazzi06} to SDSS early-type ($C>2.8$) galaxies.}\label{fig:relations_passive}
\end{figure*}

In Fig.~\ref{fig:relations_passive} we analyze the stellar populations scaling relations for quiescent and star-forming galaxies separately.
The left-hand panels show the relations for passive galaxies (black circles), while the right-hand panels the relations for star-forming
galaxies (stars color-coded according to specific SFR; empty symbols indicate galaxies with poorly constrained metallicity). The red curve and shaded region show the corresponding median
relations computed for SDSS galaxies (same sample as in Fig.~\ref{fig:relations_all})  classified according to their total specific
SFR with the same criterion used for $z=0.7$ galaxies (i.e. $\rm SFR/M_\star<1/(5t_H)$ for quiescent galaxies). The SFR used here for SDSS are total
SFR derived in \cite{Jarle04}. 

The luminosity-weighted ages (Fig.~\ref{fig:relations_passive}a) of quiescent galaxies increase
with increasing mass in a similar way as local quiescent galaxies. A linear regression fit\footnote{For all the linear relations discussed
here, we use the IDL routine {\sc mpfitexy} and we account for errors on both variables. The fitted linear relations are summarized in
Table~\ref{tab:linear_fit}. The relations plotted in Fig.~\ref{fig:relations_passive} for star-forming galaxies are those fit to the `good' data only.} gives a slope for the age--mass relation of
$z=0.7$ quiescent galaxies ($0.12\pm0.05$) consistent with that of their local counterparts. The zero-point of the relation is offset to younger ages by
0.2~dex (2.8~Gyr at $10^{11}M_\odot$) with respect to local quiescent galaxies. A similar correlation between the light-weighted age of the stellar 
populations of intermediate redshift early-type galaxies and their mass was found by \cite{ferreras09} analyzing their low-resolution HST grism spectra. The zero-point 
of their relation at a mean redshift of $\sim0.7$ is remarkably consistent with ours.

The stellar metallicities of $z=0.7$ quiescent galaxies (Fig.~\ref{fig:relations_passive}b) display a similar distribution as a function of
mass as local quiescent galaxies. Indeed, within the uncertainty, the fitted slope is consistent with the local one
($0.11\pm0.10$ compared to $0.150\pm0.003$). Moreover, the zero-point at $10^{11}M_\odot$ of $0.07\pm0.03$ is only marginally lower than the zero-point
of local quiescent galaxies ($0.109\pm0.001$), confirming the results of Fig.~\ref{fig:distr_offset}. We note
that this is also consistent with the linear relation fit to SDSS early-type galaxies selected on the basis of their concentration parameter
($C>2.8$) published in \cite{gallazzi06} (shaded area in Fig.~\ref{fig:relations_passive}). 

The stellar population properties of star-forming galaxies are shown in the right-hand panels of Fig.~\ref{fig:relations_passive}. Their light-weighted mean ages
span a larger range and follow a steeper relation with stellar mass, although much more dispersed, than quiescent galaxies (panel c). This is true both at $z=0.7$
and locally, but the zero-point for the $z=0.7$ star-forming galaxies is offset by $0.36\pm0.03$dex (2.9~Gyr) to younger ages with respect to their local
counterparts and by $0.51\pm0.03$dex (5.1~Gyr) with respect to local quiescent galaxies. Incidentally, star-forming galaxies have light-weighted mean ages younger
than quiescent galaxies, at the same $z$, by a similar amount of 2.2~Gyr at both redshifts considered. The stellar metallicities of star-forming galaxies (panel d)
increase mildly with stellar mass. The zero-point at $10^{11}M_\odot$ is $0.12\pm0.05$~dex lower than star-forming galaxies at $z=0.1$ and almost 0.2~dex lower than
quiescent galaxies at both redshifts. In particular there is a tail to metallicities lower than the local $16^{th}$ percentile of star-forming galaxies, although associated to large
uncertainties. The metallicity of star-forming galaxies thus suggests at face value the need of some chemical enrichment in this class of
galaxies between $z=0.7$ and today.

\begin{center}
\begin{deluxetable*}{lccccc}
\tablecaption{Linear fits to the stellar populations scaling relations of quiescent and star-forming galaxies shown in Fig.~\ref{fig:relations_passive}.}
\tablecolumns{6}
\tablehead{\colhead{} & \colhead{slope} & \colhead{Zp at $10^{11}M_\odot$} & \colhead{$rms_{tot}$} & \colhead{$rms_{int}$} & \colhead{$<error>$}}
\startdata
\cutinhead{$z=0.7$ quiescent}
$\log(t_r/yr)-\log(M_\star/M_\odot)$ & $0.123\pm0.05$ & $9.66\pm0.01$ & 0.06 & 0.00 & 0.07 \\
$\log(Z_\star/Z_\odot)-\log(M_\star/M_\odot)$ & $0.11\pm0.10$ & $0.07\pm0.03$ & 0.13 & 0.06 & 0.16 \\
\cutinhead{$z=0.7$ star-forming}
$\log(t_r/yr)-\log(M_\star/M_\odot)$\tablenotemark{a} & $0.42\pm0.09$ & $9.35\pm0.03$ & 0.186 & 0.121 & 0.139 \\
                                     & $0.41\pm0.11$ & $9.35\pm0.04$ & 0.195 & 0.147 & 0.127 \\
$\log(Z_\star/Z_\odot)-\log(M_\star/M_\odot)$\tablenotemark{a} & $0.42\pm0.11$ & $-0.12\pm0.05$ & 0.16 & 0.00 & 0.38 \\
                                              & $0.34\pm0.10$ & $-0.08\pm0.05$ & 0.14 & 0.00 & 0.32 \\
\cutinhead{$z=0.1$ quiescent}
$\log(t_r/yr)-\log(M_\star/M_\odot)$ & $0.101\pm0.002$ & $9.865\pm0.001$ & 0.12 & 0.07 & 0.1 \\
$\log(Z_\star/Z_\odot)-\log(M_\star/M_\odot)$ & $0.150\pm0.003$ & $0.109\pm0.001$ & 0.14 & 0.11 & 0.09 \\
\cutinhead{$z=0.1$ star-forming}
$\log(t_r/yr)-\log(M_\star/M_\odot)$ & $0.257\pm0.003$ & $9.708\pm0.001$ & 0.236 & 0.204 & 0.118 \\
$\log(Z_\star/Z_\odot)-\log(M_\star/M_\odot)$ & $0.290\pm0.003$ & $0.039\pm0.001$ & 0.24 & 0.20 & 0.15 \\
\enddata\label{tab:linear_fit}
\tablenotetext{a}{The first row is the fit obtained on the whole sample, while the second row is the fit obtained only on galaxies with at
least one the two metal-sensitive indices. This second fit is the one shown in Fig.~\ref{fig:relations_passive}c,d.}
\end{deluxetable*}
\end{center}

\section{Discussion}\label{sec:discussion}
\subsection{Evolution in metallicity}\label{sec:discussion1}
In Fig.~\ref{fig:relations_all}d,e we showed that the stellar metallicity--stellar mass relation for the population as a whole is consistent with the local one, but
offset by $-0.13\pm0.04$~dex. In Fig.~\ref{fig:relations_passive}d we showed that this offset is driven almost entirely by star-forming galaxies. While quiescent
$z=0.7$ galaxies have metallicities fully consistent with their local counterparts, the stellar metallicity of star-forming galaxies is on average $0.12\pm0.05$~dex
lower than their local counterparts at a stellar mass of $10^{11}M_\odot$. This points to some level of chemical enrichment in star-forming galaxies since $z=0.7$,
contrary to passive evolution.

The assumption of no metallicity variation in the case of passive evolution may be too simplistic. Even in the case of no chemical evolution the
light-weighted mean stellar metallicity of passively evolving {\it composite} stellar populations may vary with time for the varying light-weight of different stellar generations. When the youngest
stars fade away, the metallicity measured at $z=0$ may be lower than that at higher redshift in a closed-box system (contrary to what we see). However, if galaxies at intermediate
redshift experience star formation triggered by the infall of metal-poor gas on top of an underlying old population, the metallicity measured at $z=0.7$ would be relatively low, but
then it would rise at low $z$ once the metal-poor young stars have faded. This may occur in some galaxies, but we consider it unlikely to apply to the whole $z=0.7$ population.
Overall the difference between $z=0.7$ and $z=0.1$ in the mass-metallicity relation of both the population as a whole and of star-forming galaxies only suggests that at least a
fraction of massive star-forming galaxies experience chemical enrichment since $z=0.7$.

Interestingly, a similar amount of evolution ($\sim0.1$dex) has been measured for the gas-phase oxygen abundance of star-forming galaxies by
\cite{moustakas11} from a consistent analysis of SDSS and AGES data at $z=0.05-0.75$. Similarly to their study \citep[but in contrast to other previous
works on the evolution of the gas-phase metallicity, e.g.][]{savaglio05}, we do not detect any significant variation in the {\it shape} of the
mass--stellar metallicity relation between $z=0.7$ and today both for the population as a whole and for star-forming galaxies only. While our
data indicate a steepening of the mass--metallicity relation at $M_\star<10^{10.7}M_\odot$, we caution that this result critically depends on
a handful of low-luminosity galaxies for which we cannot measure Mg-Fe composite indices and whose low metallicities would need to be confirmed.

\cite{PS13} use the average
SFHs derived by \cite{Leitner12} as a function of mass (based on \cite{karim11} stellar masses and SFRs between $z=0$ and $z=2.5$) and the non-evolving relation between
stellar mass, gas-phase metallicity and SFR \citep{Mannucci10} to predict the stellar metallicity distribution in galaxies of different mass and the evolution with
redshift of the stellar metallicity--stellar mass relation. By construction, their predictions are strictly valid only for star-forming galaxies. They find a stellar
mass--stellar metallicity relation at $z=0$ qualitatively consistent in shape with the one derived in SDSS for the sample as a whole \citep{gallazzi05} and for
star-forming galaxies shown in Fig.~\ref{fig:relations_passive}d, however shifted to higher values with respect to the observed relations. Their estimates are tied to
the gas-phase oxygen abundance as derived in \citet{Tremonti04}, thus trace directly the $\alpha$ elements, while the stellar metallicities estimated from galaxy spectra
represent the abundance of all elements heavier than hydrogen\footnote{The metallicities derived from spectra are tied to the [Fe/H] scale of the solar-scaled
\citet{bc03} SPS models. Since we have been cautious to use absorption features minimally sensitive to [$\alpha$/Fe], the derived metallicities are indeed representative
of the abundance of all elements, but a bias may be present at {\it low} metallicities where the low-[Fe/H] stars in the STELIB library have $\rm [\alpha/Fe] \neq 0$.}.
\cite{PS13} predict that since $z=0.7$ the average stellar oxygen abundance should evolve by 0.04, 0.07, 0.1~dex at
$10^{11.5}, 10^{11}, 10^{10.5}M_\odot$. This is slightly lower than what we observe (but within our uncertainties). The difference could derive from a discrepancy
between the metallicity measured from integrated galaxy spectra and their inferred metallicity distribution within galaxies assuming some average SFHs. Moreover, their method only 
describes galaxies that are still forming stars today and does not account for star-forming galaxies at $z=0.7$ that
cease their star formation.

\subsection{Evolution in mean stellar age}\label{sec:discussion2}
In Fig.~\ref{fig:relations_all}b and Fig.~\ref{fig:distr_offset} we showed that passive evolution of the entire population of $z=0.7$ galaxies would not correctly
predict the local relation for the global galaxy population even at masses greater than $10^{11}M_\odot$. Would passive evolution reproduce the present-day scaling
relations of {\it quiescent} galaxies?

\begin{figure*}
\epsscale{1.}
\plotone{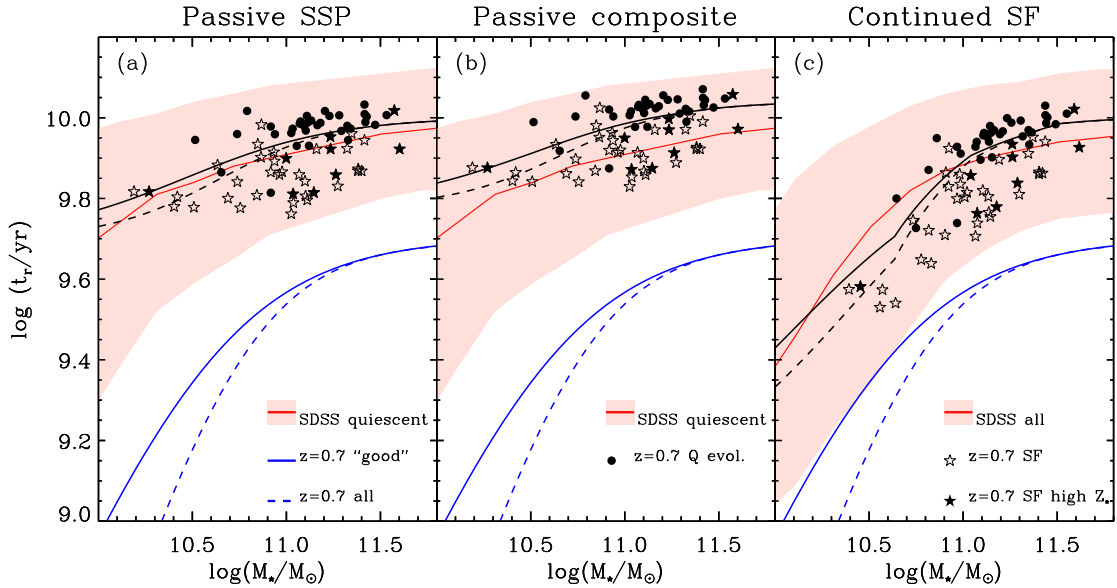}
\caption{Evolution of light-weighted age versus stellar mass from $z=0.7$ to $z=0.1$ predicted by simple passive evolution (panel a), passive evolution of
composite stellar populations (panel b) or continued star-formation assuming the average SFHs of \cite{behroozi13} for halo masses of
$\log(M_{halo}/M_\odot)=$12, 13, 14, 15 (corresponding on average to stellar masses of $\log(M_\star/M_\odot)=$10.5, 11, 11.3, 11.5; panel c). The red curve and shaded region
in each panel shows the relation for SDSS quiescent galaxies (panels a and b) and for the whole SDSS sample (panel c). The blue solid (dashed) lines indicate the
relation fit to our $z=0.7$ `good' (all) data, while the corresponding black lines show the same relation evolved to $z=0.1$. Filled circles show evolved quiescent
galaxies, while stars represent evolved star-forming galaxies (those with stellar metallicity comparable to that of local quiescent galaxies are indicated by filled
stars).}\label{fig:age_mstar_behroozi}
\end{figure*}

Fig.~\ref{fig:age_mstar_behroozi}a compares the $z=0.1$ age--mass relation of quiescent galaxies (red shaded region) with that obtained by evolving all the galaxies
in our $z=0.7$ sample by the look-back time. The ages of $z=0.7$ quiescent galaxies (filled circles) are consistent with the local quiescent population under passive
evolution. However, they display a smaller scatter in age than local galaxies and would populate only the oldest portion of present-day quiescent galaxies. As shown
in Fig.~\ref{fig:relations_passive}b their stellar metallicities are fully consistent with those of local quiescent galaxies. There is thus the need to increase the
scatter to lower ages without altering the metallicity distribution.

A modest amount of continued star-formation or `frosting' in massive quiescent galaxies could be allowed even at masses $>10^{11}M_\odot$,
which would lower their luminosity-weighted ages without affecting their (mass-weighted) metallicities \citep{trager2000,ST07}. If `frosting' rather than impulsive
starbursts is the dominant mechanism that slows down the aging of massive galaxies, then this would also explain the low fraction of
post-starburst galaxies in our sample (3\%). 
Similarly low fractions of E+A galaxies have been reported in massive field galaxies at $0.6<z<1.2$ by \cite{leborgne06} and at
$0.4<z<0.8$ by \cite{poggianti09}, with negligible evolution toward lower redshifts. Significantly higher fractions of $\sim$20\% are only found in 
high velocity dispersion clusters at intermediate redshift \citep{poggianti09}, thus hinting at possible environmental effects. 

We checked the HST V- and z-band images from GEMS of the post-starburst candidates in our sample and they do not show any sign of interaction or disturbance, but
rather their morphology appears like that of early-type spirals or S0s. This suggests that the recent quenching of star formation is not associated to a violent
event or that the signs of such an event have already faded away. These galaxies have masses close to or above $10^{11}M_\odot$, solar or super-solar metallicities
and ages between 1.5 and 3.5 Gyr. The relatively old luminosity-weighted ages may indicate that the last episode of star formation has not been very intense and that
these systems should be regarded as post-star-forming rather than post-starburst galaxies. This suggests that the triggering of a significant episode of star
formation prior to quenching does not occur frequently.

Could quenching of massive star-forming galaxies reproduce the local quiescent galaxies scaling relations? The stars in Fig.~\ref{fig:age_mstar_behroozi}a show the
$z=0.7$ star-forming galaxies passively evolved to $z=0.1$ (filled stars highlight those with a stellar metallicity in the range of local quiescent galaxies). This
shows that the younger portion of present-day quiescent galaxies could be populated by $z=0.7$ star-forming galaxies whose star formation gets truncated at $z=0.7$.
In particular such a scenario could apply to those star-forming galaxies with a metallicity already comparable to that of local quiescent galaxies.

So far we have adopted a very simplistic model of passive evolution, i.e. the aging of simple coeval stellar populations. However, 
similar considerations hold if we consider a more realistic scenario of passive evolution, i.e. the passive evolution of composite stellar populations
(Fig.~\ref{fig:age_mstar_behroozi}b). For this we consider mass-dependent average SFHs and truncate them at $z=0.7$. 

Recently \cite{behroozi13} have estimated the average SFH of galaxies in a range of halo masses from constraints to the stellar mass function,
the cosmic star formation history and the specific SFR--mass relation as a function of redshift. Their derived SFHs are fairly consistent with those obtained by
\cite{Leitner12} from constraints to the SFR--stellar mass relation as a function of redshift, which by construction include at each epoch only star forming galaxies
\citep[as opposed to][who consider all galaxy types]{behroozi13}. We model the spectral evolution of galaxies from $z=0.7$ to $z=0.1$ by applying to the BC03 SPS
models  the \cite{behroozi13} SFHs for halos of $10^{12}$, $10^{13}$, $10^{14}$, $10^{15} M_\odot$ (roughly corresponding to central galaxy stellar masses of
$10^{10.5}$, $10^{11}$, $10^{11.3}$, $10^{11.5} M_\odot$). We
notice that both at $z=0.7$ and $z=0.1$ the true r-band light-weighted ages associated with the \cite{behroozi13} SFHs are younger than those observed.\footnote{We checked that
this discrepancy is not due to a mismatch between the true light-weighted ages and those measured from the modeled spectra. For each \cite{behroozi13} SFH we produced 100
mock spectra perturbed according to the typical observational error and we fitted their absorption indices as we do for the observations. We find good agreement
between the recovered and the true light-weighted ages.} It is beyond the scope of this paper to investigate the origin of this discrepancy. We only notice that
possible reasons could be a disagreement in the average SFHs and/or the effect of differential dust attenuation on old/young stars, which is not included in
translating the \cite{behroozi13} SFH into synthetic spectra, but may indeed affect observed galaxies.

We thus consider the {\it relative} ages and we apply to the observed relation the predicted age evolution. 
By truncating these SFHs at $z=0.7$, we find the
evolution in $r$-band light-weighted age to be 6 Gyr almost independently of mass. This is larger than the passive evolution of an SSP and the predicted age--mass
relation of quiescent galaxies would be slightly older than that observed (Fig.~\ref{fig:age_mstar_behroozi}b).

Finally, we consider more complex scenarios than passive evolution and check what would be the expected evolution in light-weighted age according to the continued SFHs 
of \cite{behroozi13} as a function of mass. In this case we apply also the associated increase in stellar mass. 
The results are shown in Fig.~\ref{fig:age_mstar_behroozi}c where the $z=0.7$ evolved galaxies are compared to the age--mass relation for {\it all}
galaxies at $z=0.1$. The predicted relation would be fairly consistent with the observed one, albeit steeper. This discrepancy could be due to the simplistic
description of SFHs, intrinsic in the use of population averages, and/or to the incompleteness at low masses in our sample.

\subsection{Comparison with previous work: Does the evolution depend on environment?}\label{sec:discussion3}
In Fig.~\ref{fig:relations_passive}b we showed that quiescent galaxies have already reached the same level of enrichment as local quiescent galaxies, and may thus
simply evolve passively. The passively evolved $z=0.7$ quiescent galaxies (filled circles in Fig.~\ref{fig:age_mstar_behroozi}a) are indeed consistent with the location of present-day early-type, quiescent galaxies.

This result is consistent with a few previous studies of intermediate-redshift quiescent galaxies. \cite{kelson01} favored a scenario of
passive evolution and high formation redshift (or the redshift at which the last major star formation episode occurred) based on the analysis
of the Balmer absorption line strengths of E/S0 galaxies in four clusters at redshift between $z=0.06$ and $z=0.83$. Passive evolution of massive ($\sigma_V>175$~km/s)
red-sequence galaxies is also supported by the analysis of \cite{SB09} of 24 clusters and groups at redshift $0.45-0.75$. Similar results have
been recently obtained by \cite{Jorgensen13} from the analysis of the absorption indices versus velocity dispersion of massive early-type
galaxies in three clusters at redshift between 0.5 and 0.9, which imply a formation redshift $z\sim2.8$, consistent with the light-weighted
ages we estimate. Interestingly, their analysis of the Fundamental Plane would indicate lower formation redshifts of $\sim1.9$, under the
hypothesis of passive evolution, in disagreement with the light-weighted ages of $>10^{11}M_\odot$ quiescent galaxies.

However, our analysis also shows that the scatter in age of $z=0.7$ quiescent galaxies is too low (we in fact do
not detect any intrinsic scatter) to reproduce the local distribution. Moreover they would only populate the region between the median and the $84^{th}$ percentile of the
local population. 
Locally, the scatter in age is mass dependent, being very small or undetected at masses above $10^{11}M_\odot$ and increasing to lower masses. In \cite{gallazzi06} the
observed scatter in age for high-concentration ($C>2.8$) galaxies with masses above $10^{11}M_\odot$ was found to be comparable to their typical error on age. \cite{graves09a} from the
analysis of co-added spectra of SDSS early-type galaxies detected a range in ages at fixed velocity dispersion larger than expected from their very low formal errors
(typically 0.03~dex), which appear to be correlated with differences in surface brightness and inversely correlated with differences in metallicity \citep{graves09b}. If
there is a small intrinsic scatter in the ages of massive quiescent galaxies today, passive evolution of $z=0.7$ quiescent galaxies can be accommodated but would need to
be accompanied by additional quenching of a fraction of massive star-forming $z=0.7$ galaxies with metallicities similar to those of local quiescent galaxies, in order
to increase the scatter in age but not in metallicity. Many $z=0.7$ star-forming galaxies have indeed a metallicity comparable to that of local galaxies and they have
thus the right properties to populate, by evolving passively, the region between the median and the $16^{th}$ percentile of the age--mass relation of local quiescent
galaxies (filled stars in Fig.~\ref{fig:age_mstar_behroozi}a) without altering their metallicity distribution. 
However, passive evolution of {\it all} the massive
$z=0.7$ galaxies, besides being unlikely, would not correctly reproduce the age--mass and metallicity--mass relations for the entire galaxy population at $z=0$ nor the
metallicity--mass relation of $z=0$ star-forming galaxies. 

For field galaxies \cite{Schiavon06} argues that pure passive evolution does not apply to the population of red, emission-line free galaxies
at redshift $0.75<z<1$ in the DEEP2 survey, based on the comparison of their Balmer lines with those of the \cite{eisenstein03} coadded
spectrum of red-sequence galaxies. This could be due to either continued star formation in individual red-sequence galaxies or
the addition to the red sequence of newly quenched galaxies, or a combination of both. 
Studies of the number density evolution indicate that addition on the red sequence of newly-quenched galaxies needs to occur. This is also compatible with our analysis of quiescent and star-forming
galaxies.
In the field this should happen also at masses $>10^{11}M_\odot$ according to our
analysis \citep[or
$\sigma_V>200$km/s according to][]{Schiavon06}. This is opposed to cluster studies, in which no need for an
additional recently-quenched population is found. This suggests an
environmental dependence of the quenching epoch, in the sense that massive galaxies in clusters were quenched at an earlier epoch either after infall on the cluster or even
before, a scenario that could apply in particular to massive galaxies as suggested by recent works \citep[see e.g.][]{tinker10,wetzel13}.

\section{Summary and conclusions}\label{sec:conclusion}
We have gathered medium-resolution, rest-frame optical spectroscopic data with IMACS on Magellan for a magnitude- and mass-selected sample of
$\sim70$ galaxies at $z\sim0.7$. The quality of the data allows reliable measurements of stellar velocity dispersion and of both age- and
metallicity-sensitive absorption features in about 85\% of the sample (for the remaining fraction only age-sensitive absorption features and
more uncertain velocity dispersion measures are
available). We apply a Bayesian statistical approach, originally developed for the
analysis of low-redshift SDSS galaxies, to interpret the strength of a set of absorption features in terms of stellar metallicity,
light-weighted age and stellar mass by means of comparison with a comprehensive library of SFHs and metallicities, based on the \cite{bc03}
SPS models. We adopt the same prior distribution of models as used for SDSS spectra in order to minimize the impact of systematic effects on
the inferred evolution between $z=0.7$ and $z=0$. Stellar masses and light-weighted ages are constrained to better than 0.2~dex and to 0.11~dex and
0.13~dex on average respectively, while the average
uncertainty on stellar metallicity is 0.3~dex for galaxies with at least one of \mgtwofe~and \mgfep~(86\% of the sample). Our $z=0.7$ sample,
which includes both quiescent and star-forming galaxies, span a range in stellar mass from about $3\times10^{10}$ to $5\times10^{11}M_\odot$
corresponding to the massive end of the SDSS distribution, a range in light-weighted age from 600~Myr to 6~Gyr and a range in metallicity similar to the
SDSS distribution from about 20\% solar to $2\times Z_\odot$. 

We explore the distribution of $z=0.7$ massive galaxies in stellar metallicity and light-weighted age as a function of stellar mass. We find
that the population as a whole follows stellar metallicity--stellar mass and stellar age--stellar mass relations qualitatively similar to the local
relations with both age and metallicity increasing with mass and showing a flattening of the trend above $10^{11}M_\odot$. The parameters
describing the {\it shape} of the relations are consistent within the uncertainties with those fitted to low-redshift galaxies, with only a
tentative evidence for an increase by 0.3~dex in the characteristic mass (where the relation starts to flatten) of the age--mass
relation with respect to the local value. The zero-point and the shape of the $z=0.7$ relations are however inconsistent with pure passive
evolution of the population as a whole: {\it i)} the age--mass relation at $z=0.7$ is shifted by only $-0.28$~dex ($\sim4$Gyr, i.e. less than
the time elapsed between $z=0.7$ and $z=0.1$) from the local relation
and is too shallow with respect to the passively-evolved $z=0.1$ relation, implying a mass-dependent evolution in age that is slower than a
purely passive evolution; {\it ii)}
the median metallicity at $10^{11.5}M_\odot$ is lower by 0.13~dex than in the local sample. These results indicate that
at least a fraction of $z=0.7$ massive ($>3\times10^{10}M_\odot$) galaxies have to experience some level of star formation and increase their stellar metallicity in the
last 5~Gyr.

We considered more complex scenarios than pure passive evolution by adopting the average SFHs for four different stellar masses estimated by \cite{behroozi13} including
galaxies with any star formation activity. The mass-dependent relative age offset in the mass--age relation that arises from these SFHs in
the time lapse between $z=0.7$ and $z=0.1$ is consistent with the observed one. However we note that these SFHs result in a systematic absolute age offset
between the predicted and observed light-weighted ages at both redshifts.

By differentiating galaxies on the basis of their star formation activity we show that the location of $z=0.7$ galaxies in the
age/metallicity versus mass plane depends on the galaxy specific SFR. Quiescent galaxies are preferentially located at higher masses, with
72\% of them having a mass larger than $10^{11}M_\odot$, light-weighted ages typically between $\sim3$Gyr and $\sim6$Gyr, and stellar
metallicities around solar. Star-forming galaxies have a more uniform distribution in stellar mass, from $\sim10^{10}$ to $3\times10^{11}M_\odot$, they
are located on average at younger ages by 2.2~Gyr and lower stellar metallicities by 0.15~dex than quiescent galaxies, with a larger
dispersion in both parameters.

Quiescent galaxies at $z=0.7$ have metallicities consistent with those of local quiescent galaxies, thus not requiring an additional metal
enrichment toward $z=0$ in accord with what expected from passive evolution. They are on average $\sim3$Gyr younger than the present-day population
of $M_\star>10^{10}M_\odot$ galaxies. Their passively evolved ages are also consistent with the {\it range} of ages of
present-day quiescent galaxies.

However, our analysis also shows that, if evolved passively, $z=0.7$ quiescent galaxies would evolve in a
present-day population with smaller scatter and higher average age than observed in massive present-day galaxies, contributing only to the
oldest portion of the present-day populations. There is thus the need of an additional evolutionary path in order to populate the younger
portion of present-day quiescent galaxies without significantly altering the metallicity distribution. 

Studies of the redshift evolution of the number and mass density of quiescent and star-forming galaxies show that the population of
star-forming galaxies has remained constant or has even declined since $z<1$ until the present \citep{moustakas13,Brammer11,Ilbert10,Ilbert13}
with the largest decrease affecting the more actively star-forming galaxies \citep{Ilbert10}. Suppression of star formation thus contributes
to the build-up of the massive quiescent galaxy population. Newly-quenched, larger galaxies can also in part explain the redshift evolution of the zero-point 
of the mass-size relation of quiescent galaxies \citep[e.g.][]{vdW09,carollo13,krogager13,poggianti13}.

We show that massive star-forming galaxies at $z=0.7$ are on average $\sim5$Gyr younger than local quiescent galaxies, i.e. close to the expected
offset in the hypothesis of passive evolution. The most metal-rich $z=0.7$ star-forming galaxies, which have metallicities comparable to local quiescent
galaxies, have the right physical parameter values to supply the necessary population for the build-up of the red-sequence at masses
$>10^{11}M_\odot$. These comprise about 40\% of the star-forming galaxies in our sample at these masses.

However, passive evolution of the {\it whole} population of massive star-forming galaxies is ruled out by the age/metallicity--mass relation for the
population as a whole and by the metallicity distribution of star-forming galaxies only. Indeed $z=0.7$ star-forming galaxies are offset by $-0.12\pm0.05$ dex with
respect to equally massive local star-forming galaxies, and show a tail of galaxies metal-poorer than the lower percentile of the local distribution.
This suggests that even at masses $>10^{11}M_\odot$ a sizable
fraction of galaxies at $z=0.7$ must experience chemical enrichment since then.

This work shows that the combined analysis of the stellar populations in both quiescent and star-forming galaxies at intermediate redshifts
provides important constraints to the evolution of the massive galaxy population, complementary to those obtained from the evolution of the
mass and number density of galaxy populations. Multi-object spectrographs in the red-optical and near-IR on current large telescopes, such as
M2FS on Magellan, GMOS on Gemini, KMOS on VLT, MOSFIRE on Keck, and next generation telescopes, such as E-ELT and TMT, will allow to compile more statistically
significant and complete galaxy samples, extend to lower stellar masses and to redshift $z\sim2$. These will be crucial to build a complete and coherent
picture of the evolution of galaxy populations. 
\\

\acknowledgements
We thank the referee for a constructive report. 
We are grateful to Andrew Zirm and Sune Toft for useful comments on the manuscript, to 
Michele Cappellari for suggestions on the use of pPXF, to XianZhong Zheng for help with the FIDEL catalog and to Aday Robaina for help with the
GEMS images.  
The research leading to these results has received funding from the European Union Seventh Framework Programme (FP7/2007-2013) under grant
agreement n. 267251. 
A.G. also gratefully acknowledges support from the Lundbeck Foundation during part of this project and the Dark Cosmology Centre which is
funded by the Danish National Research Foundation.
This paper includes data gathered with the 6.5 meter Magellan Telescopes located at Las Campanas Observatory, Chile.

\begin{appendix}
\section{Possible systematic effects}\label{sec:physparam_systematics}
In this appendix we mention and try to quantify the potential systematic effects affecting the physical parameter estimates
of our $z=0.7$ sample. We consider observational uncertainties associated with the set of indices used in the fit, aperture effects,
and possible systematic differences introduced by our choice of model prior.
\subsection{Effects of using a smaller number of indices}\label{sec:physparam_systematics_indx}
Our optimal set of indices is composed of the five absorption features \dn, \hb, \hdg, \mgfep~and \mgtwofe. However, for
30/73 galaxies not all of the default five indices are available. To test the effect on the derived physical parameters of
removing one or more indices from the optimal set, we have repeated the analysis on the 43 galaxies with all the indices
measured but considering different subsets of indices each time. In particular we consider the following cases:\\
1) {\it \dn, \hb, \hdg, \mgtwofe}. This case applies to 14 galaxies in the sample. All the derived physical parameters agree well with the default
estimates. Also for stellar metallicity the majority of galaxies have a negligible offset with respect to the default estimate, but 16\% of the
galaxies have an estimate smaller by 0.1~dex with respect to the fit with all five indices. Having checked, in this and in the following cases, that
those galaxies whose parameter estimates deviate from the default ones do not occupy a particular region in the index-index planes, we interpret this
as a 16\% probability that the 14 galaxies in the sample for which we use this index set have a metallicity underestimated by 0.1~dex.
We have also verified that in practice by excluding \mgfep~from the fit to all galaxies results in scaling relations fully consistent with the default ones.\\
2) {\it \dn, \hb, \hdg, \mgfep}. This case applies to only 2 galaxies in our sample. Removing \mgtwofe~we find larger differences with the
default estimates for metallicity and stellar mass: while for the majority of the galaxies we do not see any systematic difference, we find
that 30\% of the galaxies have a metallicity lower by 0.2~dex on average and 16\% of the galaxies have a stellar mass higher by 0.15~dex. We
interpret this as a 30\% probability of a 0.2~dex underestimate for stellar metallicity and a 16\% probability of a 0.15~dex overestimate in
stellar mass. \\
3) {\it \dn,\hb, \mgfep, \mgtwofe}. This case applies to only 1 galaxy. 
The only significant deviations are seen for age estimates for which we estimate a probability of 11\% to be overestimated by 0.3~dex.\\
4) {\it \dn, \hdg, \mgtwofe}. This case applies to only 1 galaxy. The results are very similar to case 1.\\
5) {\it \dn, \hdg, \mgfep}. This case applies to only 2 galaxies. Excluding \hb~does not affect significantly the parameter estimates. The
parameters correlate well with those estimated with the default complete set of indices with a small scatter. Only for  11\% and 7\% of the
galaxies metallicity tends to be 0.15~dex lower and age tends to be 0.15~dex higher, respectively.\\
6) {\it \dn, \hdg, \hb}. This case applies to 10 galaxies in the sample. Not surprisingly, in this case we do not have a strong constraint
on stellar metallicity.  As a result, while the age estimates agree with the default ones with a scatter of 0.02~dex, stellar metallicity
tends to be underestimated by 0.14~dex on average with a scatter of 0.2~dex and stellar mass tends to be underestimated by 0.06~dex with a
scatter of 0.09~dex. The 10 galaxies in our sample for which only these three indices are available are flagged in the analysis plots
since their metallicity estimates are very uncertain and may be biased low. We stress though that we do not find evidence for a bias in
their age estimates. 

Table~\ref{tab:syserr} summarizes the mean systematic uncertainties expressed as the difference in parameter estimates between the modified fit and 
the default one.
We have also checked that the indices not included in the fit are well recovered within the observational uncertainties by the best-fit model 
in each of the above cases. In particular the distributions of $\rm (I^{fit}-I^{obs})/\sigma_I$ are of similar quality as those in Fig.~\ref{fig:didx}. 
The only exceptions are i) \hdg~which tends to be underestimated in case 3, and ii) \mgtwofe~and \mgfep~which tend to be underestimated in case 6.

\begin{center}
\begin{deluxetable}{lccc}
\tablecaption{Systematic uncertainties}
\tablecolumns{4}
\tablewidth{10cm} 
\tablehead{ & \colhead{$\Delta(logM_\star)$} & \colhead{$\Delta(logZ_\star)$} & \colhead{$\Delta(logt)$}}
\startdata
idxset1  &  $0.02 (0.03)$       &   $-0.03 (0.03)$\tablenotemark{a}   &  $0.00 (0.01)$ \\
idxset2  &  $0.03 (0.08)$\tablenotemark{b}       &   $-0.07 (0.16)$\tablenotemark{c}     &  $0.00 (0.02)$ \\
idxset3  &  $0.01 (0.05)$       &   $-0.01 (0.06)$     &  $0.03 (0.10)$\tablenotemark{d} \\
idxset4  &  $0.02 (0.03)$       &   $-0.04 (0.07)$\tablenotemark{a}     &  $0.01 (0.04)$ \\
idxset5  &  $0.04 (0.07)$       &   $-0.08 (0.16)$\tablenotemark{e}      &  $0.02 (0.05)$\tablenotemark{f} \\
idxset6  &  $0.06 (0.09)$       &   $-0.14 (0.20)$     &  $0.01 (0.02)$ \\
\hline
prior I  &  $0.10 (0.07)$       &   $-0.11 (0.11)$     &  $0.11 (0.07)$ \\
prior II &  $0.00 (0.05)$       &   $-0.07 (0.08)$     &  $-0.05 (0.05)$\\
\hline
aperture &                      &   $-0.05:-0.07$\tablenotemark{g}    &  $-0.05:-0.15$\tablenotemark{g} \\ 
         &                      &   $-0.09:-0.13$\tablenotemark{h}    &  $-0.10:-0.27$\tablenotemark{h}\\ 
\enddata\label{tab:syserr}
\tablecomments{Estimate of systematic uncertainties associated to the use of different subsets of absorption features (as in Appendix~\ref{sec:physparam_systematics_indx}), 
different assumptions on the SFH (prior I: allowing formation ages older than the Universe age; prior II: different SF timescales and fraction of burst, see 
Appendix~\ref{sec:physparam_systematics_prior}), and aperture effects (see Appendix~\ref{sec:physparam_systematics_aperture}). The table summarizes the mean difference 
(and standard deviation in parenthesis) between 
the modified parameter estimate and the default one for each case.}
\tablenotetext{a}{We estimate a probability of 16\% for $\Delta(logZ_\star)$ around $-0.1$ dex.}
\tablenotetext{b}{We estimate a probability of 16\% for $\Delta(logM_\star)$ around $0.15$ dex.}
\tablenotetext{c}{We estimate a probability of 30\% for $\Delta(logZ_\star)$ around $-0.2$ dex.}
\tablenotetext{d}{We estimate a probability of 11\% for $\Delta(logt)$ around $0.3$ dex.}
\tablenotetext{e}{We estimate a probability of 11\% for $\Delta(logZ_\star$) around $-0.15$ dex.}
\tablenotetext{f}{We estimate a probability of 7\% for $\Delta(logt)$ around $0.15$ dex.}
\tablenotetext{g}{Range of aperture bias estimated for our $z=0.7$ sample.}
\tablenotetext{h}{Range of aperture bias estimated for SDSS massive galaxies.}
\end{deluxetable}
\end{center}

\subsection{Aperture effects}\label{sec:physparam_systematics_aperture}
Both our $z=0.7$ observations and the SDSS spectroscopic data are obtained through apertures of fixed size and the fraction of light sampled may be
different for $z=0.7$ and SDSS galaxies, thus potentially affecting the inferred evolution. 
For the galaxies in our $z=0.7$ sample the median effective radius in the GEMS $z$ band is 0.69$\arcsec$, which means that the 1$\arcsec$ slit covers
typically 73\% of the galaxy extent in this band. For comparison, the fraction of the $r$-band half-light
Petrosian radius of SDSS galaxies covered by the SDSS 3$\arcsec$-diameter fiber is 63\% (67\% in the $z$-band). We can also 
estimate the fraction of light sampled by the slit by comparing the I-band flux obtained integrating the
observed spectrum with the total I-band magnitude from \combo. The median ratio is 0.69. For comparison, the SDSS
fiber collects typically 30\% of the total $r$-band flux (ratio between the fiber flux and the Petrosian flux).
In \cite{Gallazzi08} we quantified how much the stellar metallicity and age estimates appear to vary as a
function of the fraction of light collected by the fiber ($\Delta f$) for galaxies with similar stellar mass and
concentration parameter. We found the largest effect to be in the stellar metallicity of high-concentration galaxies
with masses above $10^{10.5}M_\odot$ and in the light-weighted ages of galaxies more massive than
$10^{11}M_\odot$. Assuming the median fraction of the light in the fiber or slit, this translates into a
potential overestimate of the stellar metallicity of massive SDSS galaxies between 0.09~dex and 0.13~dex and an
overestimate of the light-weighted age of 0.1~dex (high concentration) or 0.27~dex (low concentration). Assuming the same stellar
population gradients in the $z=0.7$ galaxies, hence the same variation in parameter estimates with $\Delta f$, we
would quantify an overestimate of the stellar metallicity between 0.05~dex and 0.07~dex and an overestimate of
the light-weighted age between 0.05~dex and 0.15~dex, on average (see Table~\ref{tab:syserr}). A difference of the order of 0.05~dex between the SDSS
stellar metallicity estimates and the $z=0.7$ ones could thus be accounted for by the different extent of the
aperture bias on the two samples. The conclusions on the evolution of quiescent and star-forming galaxies are largely unaffected by this possible offset.
The aperture effects so estimated would also go in the direction of reducing
the difference in light-weighted age between the SDSS and the $z=0.7$ sample, making the agreement with a
passive evolution model worse. We note that these are rough conservative estimates of the potential aperture bias. 

\subsection{Model prior assumptions}\label{sec:physparam_systematics_prior}
{\it I. Effects of limiting the maximum possible age to the age of the Universe.}\\
When fitting the observed spectra at $z=0.7$ we apply to the Monte Carlo model library the physically motivated cut in formation age to be younger than the
age of the Universe at $z=0.7$. We have tested how much our results, in particular on age, are sensitive to this imposed cut, or in other words how good
are our data in constraining the ages to be younger than the Universe age, as they should be. By removing the cut in formation age and allowing ages as old
as the present-day Universe, we still find that all but five galaxies are assigned a light-weighted age younger than the Universe age at $z=0.7$. However,
we find that the derived light-weighted ages would be on average $0.11$~dex older with respect to the default fit, with a scatter of 0.07~dex. The effect
is visible in particular for the oldest and most massive galaxies. The peak of the age distribution would be closer to the age of the Universe at $z=0.7$
and only $\sim1.5$~Gyr younger than the present-day one. Stellar metallicities would be on average 0.11~dex lower, with no clear dependence on metallicity.
Stellar masses would also be larger by 0.1~dex on average with a scatter of 0.07~dex. We stress that these are rather extreme conditions. The conclusions
on the evolution of the scaling relations would be qualitatively the same, with differences going in the direction of further disfavoring passive
evolution of all galaxies.

{\it II. Effects of changing the prior in timescale and burst fraction.}\\
In order to minimize systematic effects in the inferred evolution to low redshift, we adopt the same prior on SFHs as for SDSS. However we
consider the possibility that the distribution in SFHs at intermediate redshift be different than locally. In particular we have increased the
fraction of recent bursts from 10\% to 50\% and we have allowed for shorter timescales of the exponential SFHs (namely we adopt a prior
distribution uniform in 1/$\tau$ for timescales longer than 1~Gyr and uniform in $\tau$ for timescales between 1 and 0.1~Gyr). We find that the
stellar masses are in good agreement with our default estimates with a scatter of 0.05~dex. However light-weighted ages would be 0.05~dex
younger, especially for galaxies with $\log(t/yr)>9.4$, while stellar metallicities would be 0.07~dex lower with no dependence on metallicity
itself. The general conclusions would not change significantly. In particular, 
the effect on light-weighted ages would alleviate the discrepancy between the ages of $z=0.1$ galaxies and the passively-evolved ages of $z=0.7$
galaxies, while the effect on metallicity would strengthen the result on the metallicity evolution of star-forming galaxies.\\
The mean estimated systematic uncertainties associated to these priors are summarized in Table~\ref{tab:syserr}.

\end{appendix}

\end{document}